\documentstyle[preprint,aps,eqsecnum,prd,floats,epsf,rotate]{revtex}
\def\PR  #1 #2 #3 {Phys.~Rev.~{\bf #1}, #2 (#3)}
\def\PRL #1 #2 #3 {Phys.~Rev.~Lett.~{\bf #1}, #2 (#3)}
\def\PRD #1 #2 #3 {Phys.~Rev.~D~{\bf #1}, #2 (#3)}
\def\PLB #1 #2 #3 {Phys.~Lett.~{\bf B#1}, #2 (#3)}
\def\NPB #1 #2 #3 {Nucl.~Phys.~{\bf B#1}, #2 (#3)}
\def\RMP #1 #2 #3 {Rev.~Mod.~Phys.~{\bf #1}, #2 (#3)}
\def\ZPC #1 #2 #3 {Z.~Phys.~C~{\bf #1}, #2 (#3)}

\begin{document}
\draft
\tighten
 
\rightline{hep-ph/9608280}
\rightline{TUM-HEP-236/96}
\rightline{ILL-(TH)-96-4}
\medskip
\rightline{August 1996}
\bigskip\bigskip

\begin{center} 
{\Large \bf 
Ruling Out a Strongly-Interacting \\ 
\vspace{6mm}
Standard Higgs Model} \\  
\bigskip\bigskip\bigskip\bigskip
{\large\bf K.~Riesselmann}\\ 
\medskip 
Institut f\"ur Theoretische Physik\\
Technische Universit\"at M\"unchen \\ 
James-Franck-Stra\ss e \\ 
85747 Garching b.~M\"unchen \\ Germany \\ 
\bigskip\bigskip\bigskip 
{\large\bf S.\ Willenbrock} \\ 
\medskip 
Department of Physics \\
University of Illinois \\ 
1110 West Green Street \\  
Urbana, IL\ \ 61801 \\
\bigskip 
\end{center} 

\bigskip\bigskip\bigskip

%%%%%%%%%%%%%%%%%%%%%%%%%%%%%%%%%%%%%%%%%%%%%%%%%%%%%%%%%%%%%%%%%%%%%%%%%%%%%%%

\begin{abstract}
  Previous work has suggested that perturbation theory is unreliable for Higgs-
  and Goldstone-boson scattering, at energies above the Higgs mass, for
  relatively small values of the Higgs quartic coupling $\lambda(\mu)$.  By
  performing a summation of nonlogarithmic terms, we show that perturbation
  theory is in fact reliable up to relatively large coupling.  This eliminates
  the possibility of a strongly-interacting standard Higgs model at energies
  above the Higgs mass, complementing earlier studies which excluded strong
  interactions at energies near the Higgs mass.  The summation can be
  formulated in terms of an appropriate scale in the running coupling,
  $\mu=\sqrt{s}/e\approx\sqrt{s}/2.7$, so it can easily be incorporated in
  renormalization-group improved tree-level amplitudes as well as higher-order
  calculations.
\end{abstract}

\addtolength{\baselineskip}{9pt}

\newpage

%%%%%%%%%%%%%%%%%%%%%%%%%%%%%%%%%%%%%%%%%%%%%%%%%%%%%%%%%%%%%%%%%%%%%%%%%%%%%%%

\section{Introduction}

\indent\indent The electroweak interaction is a gauge theory, with the gauge
symmetry spontaneously broken to that of electromagnetism.  A major outstanding
problem in particle physics is to discover the mechanism which breaks the
symmetry.  The simplest model of the symmetry-breaking mechanism is the
standard Higgs model, in which a fundamental scalar field acquires a
vacuum-expectation value $v = (\sqrt 2 G_F)^{-1/2}= 246$ GeV \cite{WS}.  The
particle content of the model is a spin-zero boson, dubbed the Higgs boson
($H$), and three massless Goldstone bosons ($w^+,w^-,z$) which are ultimately
absorbed by the weak gauge bosons.

It has been established that the standard Higgs model exists only up to a
cutoff energy $\Lambda$ at or before which the model must be subsumed by a more
fundamental theory~\cite{E}. Thus the standard Higgs model is regarded as an
effective field theory, valid for energies less than $\Lambda$. The maximal
allowed value of $\Lambda$ decreases with increasing Higgs mass, $m_R$.
Demanding, for consistency, that $m_R<\Lambda$ leads to an upper bound on the
Higgs mass~\cite{DN}.

In this paper we address the question of whether the standard Higgs model can
be strongly interacting at energies and Higgs masses less than the cutoff
$\Lambda$.  By ``strongly interacting'' we mean that the Higgs self-coupling,
$\lambda$, is so large that perturbation theory is unreliable.  There are two
scenarios which yield a large value for the Higgs coupling: (i) The running
coupling $\lambda(\mu)$ increases with increasing scale $\mu$, leading to a
strong coupling at energies above the Higgs mass; (ii) A Higgs
mass, $m_R$, much larger than the vacuum-expectation value, $v=246$ GeV,
results in a large coupling $\lambda(m_R)\equiv m_R^2/2v^2$.  Since the Higgs
model is constrained by the cutoff $\Lambda$, the two possibilities lead to the
following questions:
\begin{itemize}
\item Can the running coupling $\lambda(\mu)$ become strong for
  energies $\sqrt{s}$ in the range $m_R < \sqrt{s} < \Lambda$?
\item Can $\lambda(m_R)$ be strong for values of the Higgs mass
  $m_R$ below the cutoff $\Lambda$?
\end{itemize} 
The first question is related to high-energy processes such as Higgs- and
Goldstone-boson scattering.  The second question can be investigated in the
context of Higgs-boson decays.  Since both of these processes have been
calculated at next-to-next-to-leading order in perturbation theory, they are
appropriate indicators of the reliability of perturbation theory.

A popular way to model the cutoff $\Lambda$ is to use a lattice with a finite
lattice spacing $a$ \cite{DN}.  For a review of early work, we refer to
Ref.~\cite{E}, whereas a more current set of references is given in, for
example, Refs.~\cite{GKNZ,HKNV}.  Using such an approach, the cutoff $\Lambda$
is proportional to $a^{-1}$.  When lattice-spacing effects on physical
quantities are small, the model is equivalent to the standard Higgs model in
the continuum.  When lattice-spacing effects are large, the standard Higgs
model ceases to exist as an effective field theory. This observation can be
used to establish an upper bound on the Higgs mass.

Using the condition that the inverse lattice spacing be greater than twice the
Higgs mass ($a^{-1}>2m_R$), L\"uscher and Weisz \cite{LW1,LW2} determined an
upper bound on $\lambda(m_R)$ of 3.2, corresponding to an upper bound on the
Higgs mass of 630 GeV.  A subsequent study \cite{GKNZ} found a similar
upper bound on $\lambda(m_R)$. Alternative formulations of the
lattice action can increase the bound slightly \cite{GKNZ,HKNV}.
L\"uscher and Weisz argued that perturbation theory is reliable for a Higgs
coupling of $\lambda(m_R)=3.2$. They based their statement on observations
regarding three perturbative observables: (i) Such a value of $\lambda(m_R)$
yields a perturbative Higgs width which is much less than its mass, (ii)
Two-loop perturbative cross sections at threshold in the symmetric phase of the
model are apparently convergent for such a coupling, and (iii)
This coupling is less than the perturbative unitarity bound\footnote{The
  perturbative unitarity ``bound'' is not an absolute bound on the possible
  value of $\lambda$ (or the Higgs mass), but rather the value above which the
  coupling is strong.  In contrast, the lattice bound on the coupling is truly
  a bound, in the sense that the standard Higgs model cannot exist as an
  effective field theory if the coupling exceeds this value.} on $\lambda$.
They therefore concluded that there is no strongly-interacting Higgs model in
which the cutoff is substantially greater than the Higgs mass.

Recent perturbative studies of high-energy Higgs- and Goldstone-boson
scattering in the broken phase of the model have led to a different conclusion
\cite{DJL,DMR2,R,NR}.  Considering the high-energy limit, the relevant coupling
of these observables is the running coupling $\lambda(\mu)$, where $\mu$ is of
the order of the center-of-mass energy, $\sqrt{s}$.  Using a variety of 
criteria, all high-energy studies found
that the two-loop high-energy perturbative amplitudes do not converge
satisfactorily for $\lambda(\sqrt s) \gtrsim 2.0 - 2.3$.  For example, Durand,
Lopez, and Johnson argued that perturbation theory is unreliable for
$\lambda(\sqrt s)$ as low as $2.0$ \cite{DJL}.  This conclusion was based on a
one-loop analysis of partial-wave unitarity in Higgs- and Goldstone-boson
scattering, and on the lack of convergence of the perturbation series. Using a
variety of additional criteria to judge the convergence of the perturbative
series, subsequent analyses at two loops have only served to reinforce this
conclusion \cite{DMR2,R,NR}.  Since the running coupling $\lambda(\sqrt{s})$
can attain a value of $2.0 - 2.3$ for values of $\sqrt{s}<\Lambda$, the
standard Higgs model could be strongly interacting at energies above the Higgs
mass but below the cutoff $\Lambda$.

In this paper we reinvestigate the perturbative behaviour of the high-energy
Higgs- and Goldstone-boson scattering.  We introduce a summation procedure
which shifts the value of the coupling at which perturbation theory becomes
unreliable to $\lambda \approx 4.0$.  Requiring that the energy $\sqrt{s}$ be
less than the cutoff $\Lambda$, the perturbative bound $\lambda \approx 4.0$ is
large enough to ensure the absence of a strongly-interacting Higgs sector at
high energies.  Thus our summation procedure restores the convergence of
perturbation theory at energies above the Higgs mass but below the cutoff
$\Lambda$.  This is a new result, and complements the result of L\"uscher and
Weisz on the impossibility of a strong Higgs coupling at $\mu\approx m_R$.  We
conclude that the possibility of a strongly-interacting standard Higgs model
is eliminated at all energies.

Our summation procedure is based on identifying a certain class of Feynman
diagrams which can be summed by an appropriate scale $\mu$ in the running
coupling $\lambda(\mu)$.  Calculating high-energy Higgs- and Goldstone-boson
scattering, all previous analyses have implicitly or explicitly chosen
$\mu=\sqrt s$ \cite{DJL,DMR2,R}, or have varied the scale about this value
\cite{NR}.  We argue that a better scale is $\mu=\sqrt s/e\approx
\sqrt{s}/2.7$.  This scale corresponds to a summation of a universal
nonlogarithmic term which accompanies the leading logarithms in the Higgs- and
Goldstone-boson scattering diagrams.  We show that this summation greatly
improves the convergence of perturbation theory: the coefficients of the
perturbative series are greatly reduced as seen up to two loops, and the scale
dependence is significantly reduced when varying $\mu$ around $\sqrt{s}/e$
(rather than $\sqrt{s}$).

The remainder of the paper is organized as follows.  In section 2 we reanalyse
Higgs- and Goldstone-boson scattering up to two loops, and argue for the
appropriate scale $\mu$ in the running coupling $\lambda(\mu)$.  We consider
the convergence of perturbation theory and the partial-wave unitarity of these
scattering amplitudes with this improved choice of scale.  We derive the value
of the running coupling for Higgs- and Goldstone-boson scattering at the cutoff
$\Lambda$, and find that it is within the range of validity of perturbation
theory.  In section 3 we briefly review the Higgs decay amplitude at two loops.
We show that for the decay amplitude, the natural scale $\mu=m_R$ is unaffected
by our summation procedure.  The value of $\lambda(m_R) \approx 4.0$ at which
perturbation theory becomes unreliable in Higgs decays remains unchanged from a
previous analysis.  In section 4 we discuss some phenomenological consequences
of our work for scattering cross sections. We summarize our results in section
5.

\section{Higgs- and Goldstone-boson scattering}

\begin{figure}[hb]
\vspace*{13pt}
\centerline{
\epsfysize=3.5in \rotate[r]{\epsffile{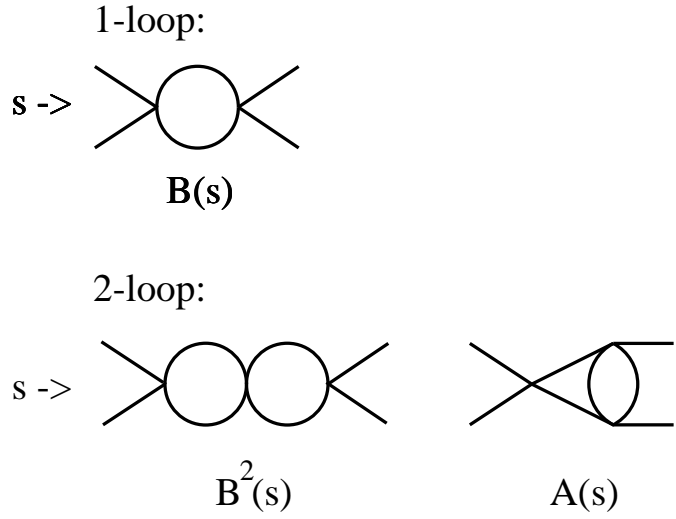}}
}
\vspace{0.2in}
\caption{
  Topologies of $s$-channel Feynman diagrams contributing to Higgs- and
  Goldstone-boson scattering at one and two loops. The $t$- and $u$-channel
  diagrams are obtained by crossing relations. }
\label{figdiag}
\end{figure}

An estimate of the value of the Higgs running coupling $\lambda(\mu)$ at which
perturbation theory becomes unreliable can be obtained from the evaluation of
$2\rightarrow 2$ scattering processes in the standard Higgs model at high
energy $(s\gg m_R^2)$ \cite{DM,LQT,DJL,DMR2,R,NR}. The basis for such analyses
is the generic high-energy scattering amplitude of Higgs and Goldstone bosons,
$a(s,t,u)$.  Up to two loops the relevant Feynman scattering diagrams are shown
in Fig.~\ref{figdiag}.  Including the combinatoric factors, the unrenormalized
scattering amplitude is
\begin{eqnarray}
a_0(s,t,u)&\stackrel{s\gg m_R^2}{=}& 
- 2 \lambda_0 
- \lambda_0^2 \,[\,(10+2n_g) B(s) + 4B(t) + 4B(u) \,]\nonumber\\
& & -\lambda_0^3 \,\left[\, (38+16n_g+2n_g^2) [B(s)]^2 + 8[B(t)]^2 + 8[B(u)]^2 
\right.\nonumber\\
& & \phantom{-\lambda_0^3 [\,} \left.+ (104+24n_g)A(s) + (56+8n_g)A(t) +
(56+8n_g)A(u)  \,\right] \nonumber\\
& & + O(\lambda_0^4) + O(m_R^2/s) \, .
\label{a02lp}
\end{eqnarray}
Here $n_g=3$ is the number of Goldstone bosons, and the quantity $\lambda_0$
denotes the bare Higgs quartic coupling.  The functions $A$ and $B$ correspond
to the Feynman diagrams depicted in Fig.~\ref{figdiag}: $B$ is the one-loop
``bubble'' diagram, and $A$ is the two-loop ``acorn'' diagram \cite{DMR1}.
The renormalized amplitude is 
\begin{equation}
a(s,t,u) = \left.\left(Z_w^{1/2}\right)^4\,
a_0(s,t,u)\right|_{\lambda_0=\lambda+\delta\lambda}\,,
\label{renamp}
\end{equation}
where $Z_w$ is the wavefunction renormalization of the Goldstone-boson fields
and $\delta\lambda$ is the coupling counterterm.  

It is standard in both lattice and continuum calculations to express the
renormalized amplitude in terms of the Higgs mass, $m_R$, and the coupling
\cite{LW1,DJL,MW}\footnote{To make contact with the notation of
  Refs.~\cite{LW1,LW2} and of most subsequent lattice work, note that $g_R
  \equiv 6\lambda(m_R)$.}
\begin{equation}
\lambda(m_R) \equiv \frac{1}{2}\frac{m_R^2}{v^2},
\label{COUPLING}
\end{equation}
where $v$ is the vacuum-expectation value of the Higgs field, defined by
$v\equiv (\sqrt 2 G_F)^{-1/2}=246$ GeV, with $G_F$ extracted from some
low-energy weak process, such as muon decay.  Up to numerically small
corrections (see Appendix \ref{mass}), $m_R$ corresponds to the physical Higgs
mass.  The wavefunction renormalization constant and the coupling counterterm
are known up to two loops \cite{DMR1,G1}.

In the limit $s\gg m_R^2$, the physical $2\rightarrow2$ scattering
amplitudes of the Higgs and Goldstone bosons are related to the generic 
high-energy amplitude $a(s,t,u)$ in the following way: 
\begin{eqnarray}
\label{wwzz1}
A_{WW\rightarrow ZZ} = \frac{Z_H}{Z_w}A_{HH\rightarrow ZZ}
= \frac{Z_H}{Z_w}A_{HH\rightarrow WW}
&=& a(s,t,u)\,,\\
\label{wwzz2}
A_{WW\rightarrow WW} 
&=& a(s,t,u) + a(t,s,u)\,,\\
\label{wwzz3}
A_{ZZ\rightarrow ZZ} = \frac{Z_H^2}{Z_w^2}A_{HH\rightarrow HH} 
&=& a(s,t,u) + a(t,s,u) + a(u,t,s)\,.
\end{eqnarray}
The wavefunction renormalization $Z_H$ of the external Higgs field is given
in\cite{DMR2,G1}.  A detailed investigation of the perturbative behaviour of
the different channels is carried out in Ref.~\cite{R}.

Of particular interest is the approximate SO(4) singlet scattering amplitude
$\tilde a^0$.  It is the $s$-wave projection of this amplitude which yields the
strongest unitarity bound in perturbation theory.  At tree level, $\tilde a^0$
equals the SO(4) singlet eigenamplitude $a^0$ considered by Lee, Quigg,
and Thacker \cite{LQT}. The corresponding tree-level SO(4) singlet eigenstate
is
\begin{eqnarray}
        \chi^0 &=& \frac{1}{\sqrt{8}}(2w^+w^-+zz+HH)\,,
\end{eqnarray}
and the tree-level eigenamplitude $a^0(\chi^0\rightarrow\chi^0)$
is expressible in terms of the generic function $a(s,t,u)$: 
\begin{equation}
a^0 = 2a(s,t,u) + \frac{3}{4}a(t,s,u) + \frac{1}{4}a(u,t,s) \,.
\end{equation}
At one loop and beyond, the eigenstate $\chi^0$ mixes with the isospin-singlet
component of the SO(4) nonet states \cite{DJL,DMR2}.  The resulting eigenstate
$\tilde\chi^0$ determines the modified eigenamplitude $\tilde a^0$. An
appropriately-normalized integral over the scattering angle yields the $J=0$
partial-wave-projected eigenamplitude $\tilde a_0^0$, the usual $s$-wave
amplitude.

Including the two-loop corrections \cite{DMR2}, we can write the
renormalization-group improved $s$-wave eigenamplitude $\tilde a_0^0$ in terms
of the running coupling $\lambda(\mu)$. Not specifying a particular choice of
$\mu$, we find:
\begin{eqnarray}
\tilde a_0^0 & = & -\left(\frac{\mu^2}{m_R^2}\right)^{\gamma}
\frac{3}{8\pi}\lambda(\mu)
\left\{1+\frac{\lambda(\mu)}{16\pi^2}
\left[12\ln\frac{s}{\mu^2}-22.27-6\pi i\right]\right.\nonumber\\
&&\phantom{-\left(\frac{\mu^2}{m_R^2}\right)^{\gamma}} +
\left(\frac{\lambda(\mu)}{16\pi^2}\right)^2\left.\left[144\ln^2\frac{s}{\mu^2}
+(-691.1-144\pi i)\ln\frac{s}{\mu^2}+1012.3+821.6i\right]\right\}\;.
\label{wwww}
\end{eqnarray}
The definition of the running coupling $\lambda(\mu)$ is given in Appendix
\ref{runcoup}.  The renormalization group is used to evolve the coupling from
the Higgs mass, see Eq.~(\ref{COUPLING}), up to a scale $\mu>m_R$.  The only
explicit dependence of the amplitude on $m_R$ occurs in the overall factor
associated with the anomalous dimension $\gamma$ of the eigenstate
$\tilde\chi^0$.  At one loop $\gamma=0$, and at two loops $\gamma$ is
numerically small \cite{DMR2}:
\begin{equation}
\gamma = -12\left(\frac{11}{2}-\pi\sqrt 3 \right)\left(\frac{\lambda}{16 
\pi^2}\right)^2 = -2.8\times 10^{-5} \lambda^2\,.
\end{equation}
Since we are concerned with values of $\lambda(\mu)<10$, we may approximate
$\gamma\approx 0$ throughout our analysis.  The eigenamplitude $\tilde a_0^0$
then has no explicit dependence on $m_R$: it depends only on the running
coupling $\lambda(\mu)$ and the scale $\mu$.  It is therefore an ideal
observable to derive perturbative upper bounds on the running coupling in the
limit $s\gg m_R^2$.

\subsection{The choice of the scale $\mu$}

The scale $\mu$ should be chosen such that the logarithms in the amplitude,
Eq.~(\ref{wwww}), are small, in order to avoid large coefficients in the
perturbative expansion.  By inspection of Eq.~(\ref{wwww}), we see that $\mu$
should be of order $\sqrt s$. This choice corresponds to a summation of the
leading logarithms into the running coupling. This observation has led to the
scale $\mu=\sqrt s$ becoming the standard in calculations of Higgs- and
Goldstone-boson scattering.  Using this scale, one finds that the perturbative
expansion of $\tilde a_0^0$ becomes unreliable for the surprisingly-low value
$\lambda(\sqrt{s})=2.0-2.3$ \cite{DJL,DMR2,R,NR}, as discussed in the
Introduction.

We argue that a more appropriate choice is $\mu=\sqrt{s}/e\approx
\sqrt{s}/2.7$.  We begin by reviewing the calculation of $a(s,t,u)$ at one
loop.  Starting from Eq.~(\ref{renamp}), the contributions to the renormalized
amplitude are the bubble diagram $B$ (Fig.~\ref{figdiag}, top), the
wavefunction renormalization $Z_w$, and counterterms:
\begin{eqnarray}
a(s,t,u) &=& - 2 \lambda(\mu)
- \lambda^2(\mu)
\left[ 16B(s) + 4B(t) + 4B(u) \right]\nonumber\\
&+& {\rm counterterms} +\; {\rm wavefunction\;renorm.}\,\,.
\label{ampstrt}
\end{eqnarray}
The one-loop renormalization-group logarithms, $\ln(p^2/\mu^2)$, arise solely
from the bubble scattering diagrams ($p^2=s,t$, or $u$), with the internal
lines of the bubble representing either Higgs- or Goldstone-boson propagators.
In the high-energy limit, $s\gg m_R^2$, the mass of the Higgs boson can be
neglected, so the evaluation of the bubble diagram $B(p^2)$
involves only massless propagators.  In dimensional regularization
($D=4-2\epsilon$), one finds at one loop
\begin{eqnarray}
B(p^2) 
&=& \frac{1}{16\pi^2} 
\left(  \Delta+ \ln\frac{\mu^2}{-p^2} + 2 
+ O(\epsilon)
\right) \,,
\label{ONELOOP}
\end{eqnarray}
where $p^2$ is the four-momentum squared flowing through the bubble, and 
\begin{equation}
\Delta\equiv\frac{1}{\epsilon}-\gamma+\ln 4\pi\;
\end{equation}
is divergent in four dimensions ($\epsilon = 0$).
The Euler constant is denoted by $\gamma$.

Evaluating the renormalized amplitude $a(s,t,u)$ according to
Eq.~(\ref{ampstrt}), all the $\Delta$'s are cancelled by the counterterms,
yielding a finite result.  However, the constant 2 appearing in
Eq.~(\ref{ONELOOP}) is {\it not} cancelled in the renormalized amplitude.  This
is because the counterterms are calculated at ``low'' energies: $p^2=m_R^2$ in
the case of Higgs counterterms, $p^2=0$ for Goldstone-boson quantities.  Hence
one cannot neglect the Higgs mass in the calculation of the counterterms, so
the counterterms do not involve massless bubble diagrams (except for those
which involve two massless Goldstone bosons).  As a result, the constant which
accompanies the divergence $\Delta$ varies from diagram to diagram when
calculating the low-energy counterterms (see Appendix \ref{massbub}), unlike
the universal constant 2 which appears in {\it all} high-energy scattering
bubble diagrams.

Putting together the various contributions, the renormalized one-loop amplitude
may be written 
\begin{eqnarray}
a(s,t,u)= -2\lambda(\mu) + \frac{\lambda^2(\mu)}{16\pi^2}\left[
\beta_0\left(\ln\frac{\mu^2}{s} + 2\right) + (10+2n_g)i\pi
- 4\ln\frac{\sin^2\theta}{4} - 4n_g - 1.35 \right]\,.
\label{ngamp}
\end{eqnarray}
The coefficient of the logarithm, $\ln(\mu^2/s)$, is the one-loop beta-function
coefficient, $\beta_0=18+2n_g=24$, and we maintain the association of the
constant 2 with the logarithm as suggested by Eq.~(\ref{ONELOOP}).  The other
terms appearing in (\ref{ngamp}) have the following origin: the imaginary part
is from the $s$-channel bubble diagrams, the angular dependence is from the
$t$- and $u$-channel diagrams (where $t,u= -s(1\pm\cos \theta)/2$, and $\theta$
is the center-of-mass scattering angle), and the term $-4n_g-1.35$ originates
from counterterms and wavefunction renormalization.  The crossed amplitude,
$a(t,s,u)$, has the same $\beta_0$ term as $a(s,t,u)$, but its imaginary part
as well as its angular dependence are different.  This leads to the universal
appearance of the term $\beta_0(\ln(\mu^2/s)+2)$ in all high-energy amplitudes
listed in Eqs.~(\ref{wwzz1})--(\ref{wwzz3}).

The two-loop Feynman scattering diagrams which contribute to the amplitude are
shown in Fig.~\ref{figdiag} (bottom row), and their analytical results are
given in Appendix \ref{scatt}.  There are two different topologies: a chain of
two bubbles, $[B(p^2)]^2$, and the ``acorn'' diagram, $A(p^2)$, which consists
of a bubble subdiagram inserted at a vertex of a bubble diagram.  Each class of
diagrams contributes to the leading logarithm at two loops, $\ln^2(\mu^2/s)$.
The chain of two bubbles clearly has a 2 associated with each logarithm since
it is the square of the one-loop bubble diagram.  According to
Eq.~(\ref{a02lp}) and taking $n_g=3$, roughly half of the two-loop leading
logarithms come from the chain of two bubble diagrams, $B^2(p^2)$.
%This class of diagrams supports the scale $\mu=\sqrt s/e$, since it eliminates
%the constant 2 associated with each logarithm in the bubble diagram. 
The other half of the two-loop leading logarithms comes from the acorn diagram,
$A(p^2)$, which is more subtle: the bubble subdiagram of $A(p^2)$ does have a
2, but the energy scale appearing in its logarithm is not $s$, but rather is an
integration variable which is integrated over when the subdiagram is inserted
into the full two-loop diagram $A(p^2)$.  The remaining second loop integration
then becomes a modified bubble diagram, with a momentum-dependent vertex (due
to the bubble subdiagram).  This loop integration does not simply yield a
constant 2.

The situation is similar at three loops and beyond.  At $n$ loops there is
always
a topology which is a product of $n$ bubbles. This class of diagrams has the
maximal number of 2's connected with the leading logarithm. Next there are
topologies which have $n-1$ bubbles, with the final integration being a
modified bubble integral.  Then there are topologies with $n-2$ bubbles, and so
on.  Starting at three loops, there also exist nonplanar graphs which cannot be
naturally viewed as being constructed from bubble graphs.  Yet their weight is
expected to be small compared to the numerous bubble related contributions to
the complete set of $n$-loop diagrams.

The universality of the term $\beta_0(\ln(\mu^2/s)+2)$ suggests that the scale
$\mu$ should be chosen to eliminate both the logarithm and the constant.  Hence
we advocate
\begin{equation}
\mu=\frac{\sqrt s}{e}\approx\frac{\sqrt s}{2.7}\,,
\end{equation}
in contrast to the usual choice $\mu=\sqrt{s}$.  Our choice of scale amounts to
summing the constant 2 along with the leading logarithm to all orders in
perturbation theory. At two loops and beyond, the new scale also reduces the
finite contributions coming from the $O(\epsilon^n)$ terms; see Appendix
\ref{resum}.   Since none of the other terms in Eq.~(\ref{ngamp}) are
proportional to $\beta_0$, it is inappropriate to choose the
renormalization-group scale $\mu$ to sum any of them.

\subsection{Testing the new scale}

A concern is that the scale $\mu=\sqrt{s}/e$ may not be an appropriate choice
for the next-to-leading logarithms, which first appear at two loops.  A ``bad''
choice of scale could result in a large two-loop coefficients.  To investigate
this aspect, we compare the perturbative expansions of the eigenamplitude
$\tilde a_0^0$ at next-to-next-to-leading order, using the scales $\mu=\sqrt s$
and $\mu=\sqrt s/e$.  Choosing the scale $\mu=\sqrt s$, one obtains from
Eq.~(\ref{wwww}) (approximating $\gamma=0$)
\begin{eqnarray}
   \tilde a_0^0 & = & -\frac{3}{8\pi}\lambda(\sqrt s)
\left\{1+\frac{\lambda(\sqrt s)}{16\pi^2}
\left[\,-22.27-6\pi i\,\right]
+\left(\frac{\lambda(\sqrt s)}{16\pi^2}\right)^2
\left[\,1012.3+821.6i\,\right]\right\}\;,
\label{wwwwconv}
\end{eqnarray}
where $\lambda(\sqrt s)$ is the three-loop running coupling evaluated at
$\mu=\sqrt s$.  The new scale $\mu=\sqrt s/e$ yields
\begin{eqnarray}
\tilde a_0^0 & = & -
\frac{3}{8\pi}\lambda(\sqrt s/e)
\left\{1+\frac{\lambda(\sqrt s/e)}{16\pi^2}
\left[\,1.73-6\pi i\,\right]
+\left(\frac{\lambda(\sqrt s/e)}{16\pi^2}\right)^2
\left[\,206.1-83.1i\,\right]\right\}\;,
\label{wwwwour}
\end{eqnarray}
where the three-loop running coupling is evaluated at $\mu=\sqrt s/e$.  We find
that the summation of the 2's greatly reduces the size of the coefficients of
the perturbative amplitude.  Furthermore, the value of the running coupling at
$\mu=\sqrt s/e$ is {\it less} than at $\mu=\sqrt s$, leading to a further
improvement in the convergence of perturbation theory.  The above results
support the improved scale at the leading-log level and also suggest that it is
the appropriate scale at the subleading level.

\subsection{Upper perturbative bound on the running coupling}

We now attempt to quantify the value of $\lambda(\sqrt s/e)$ at which 
perturbation theory becomes unreliable.  There are three criteria we can 
use to judge the convergence of perturbation theory:
(i) The size of the radiative
corrections should be small; 
(ii) The scale dependence should decrease with increasing order in 
perturbation theory; 
(iii) The amplitude should not violate perturbative unitarity by a large
amount. 

\begin{figure}[tb]
\vspace*{13pt}
\centerline{
\epsfysize=2.2in \rotate[l]{\epsffile{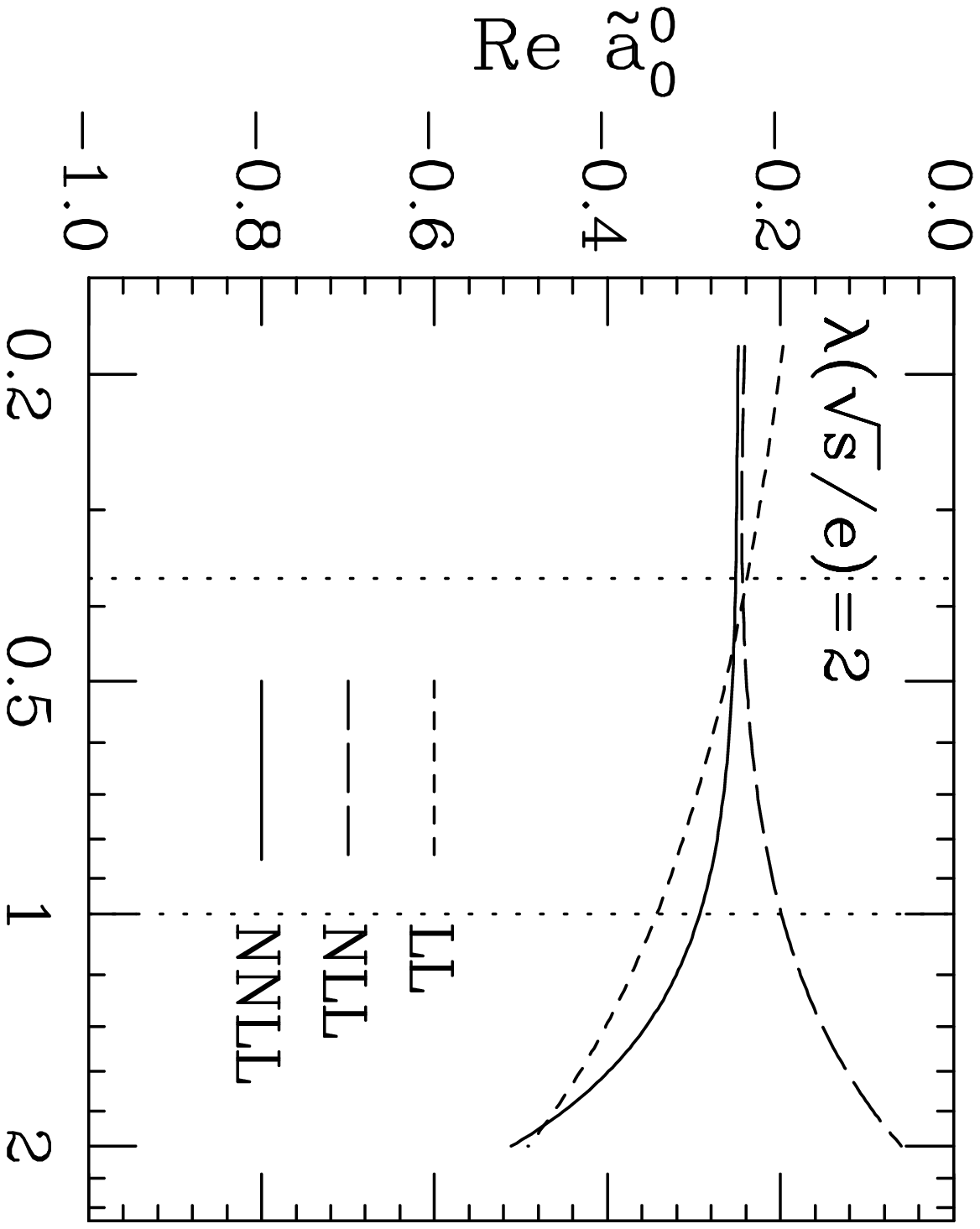}}
\hspace{1cm}\epsfysize=2.2in \rotate[l]{\epsffile{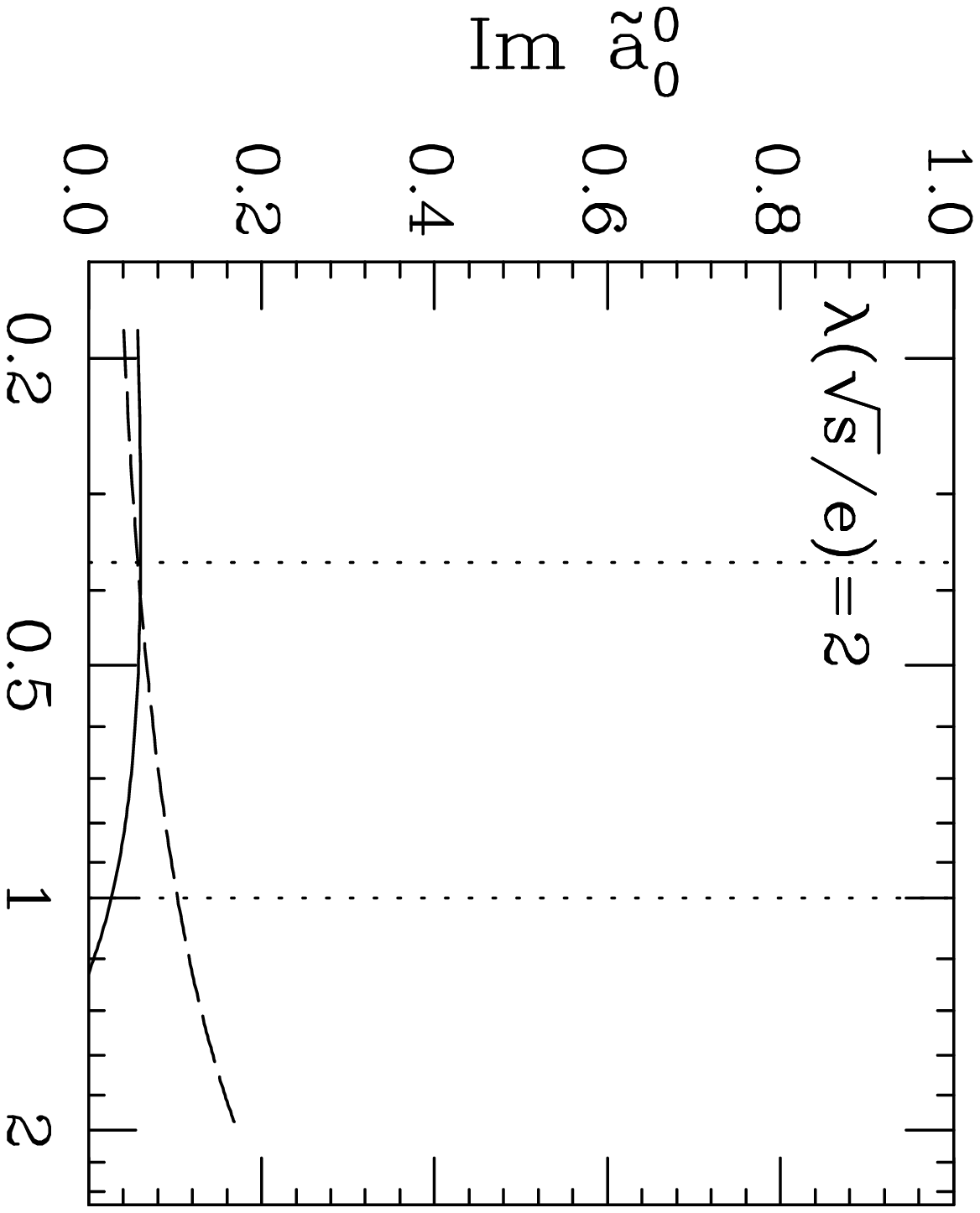}}
}
\centerline{
\epsfysize=2.2in \rotate[l]{\epsffile{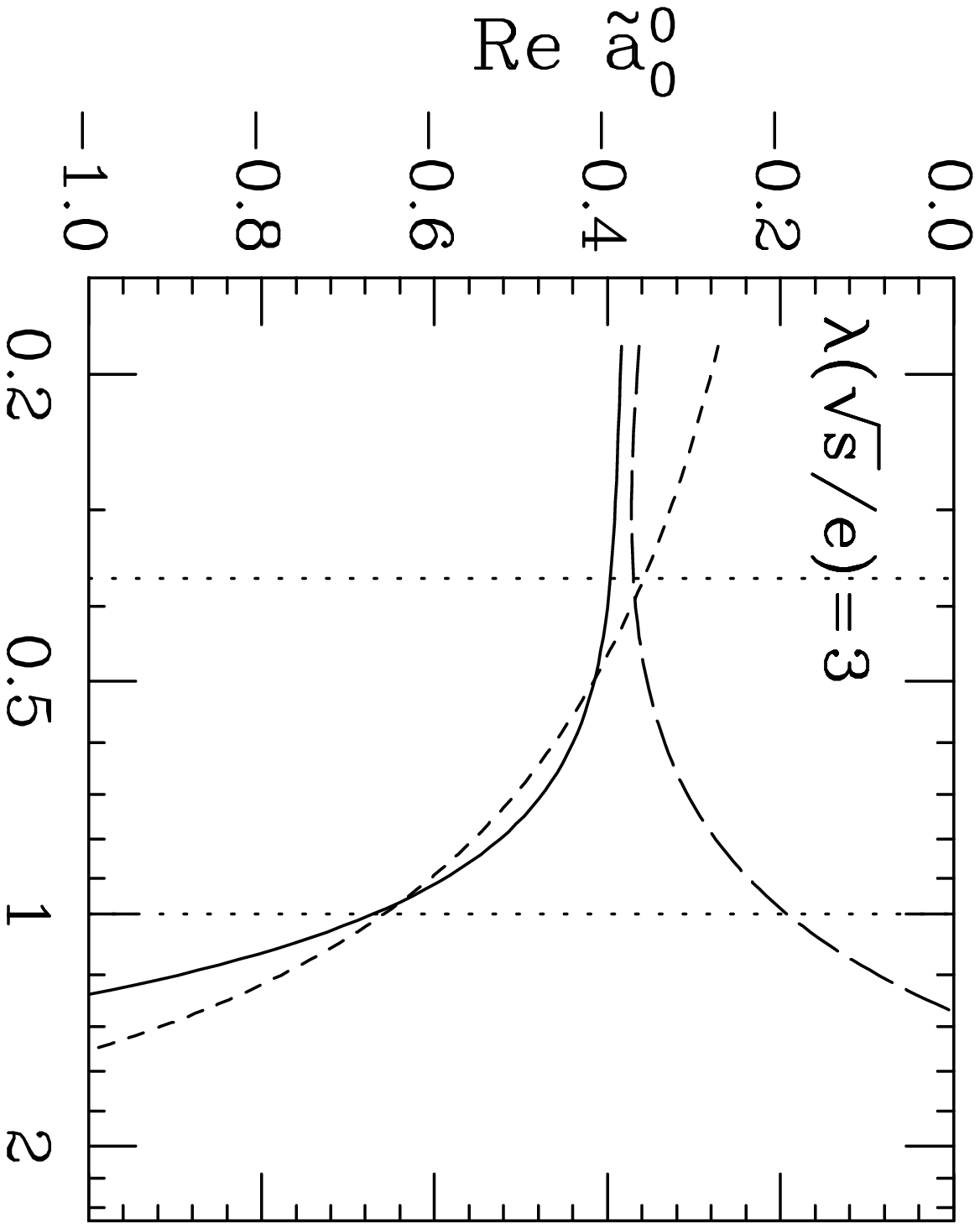}}
\hspace{1cm}\epsfysize=2.2in \rotate[l]{\epsffile{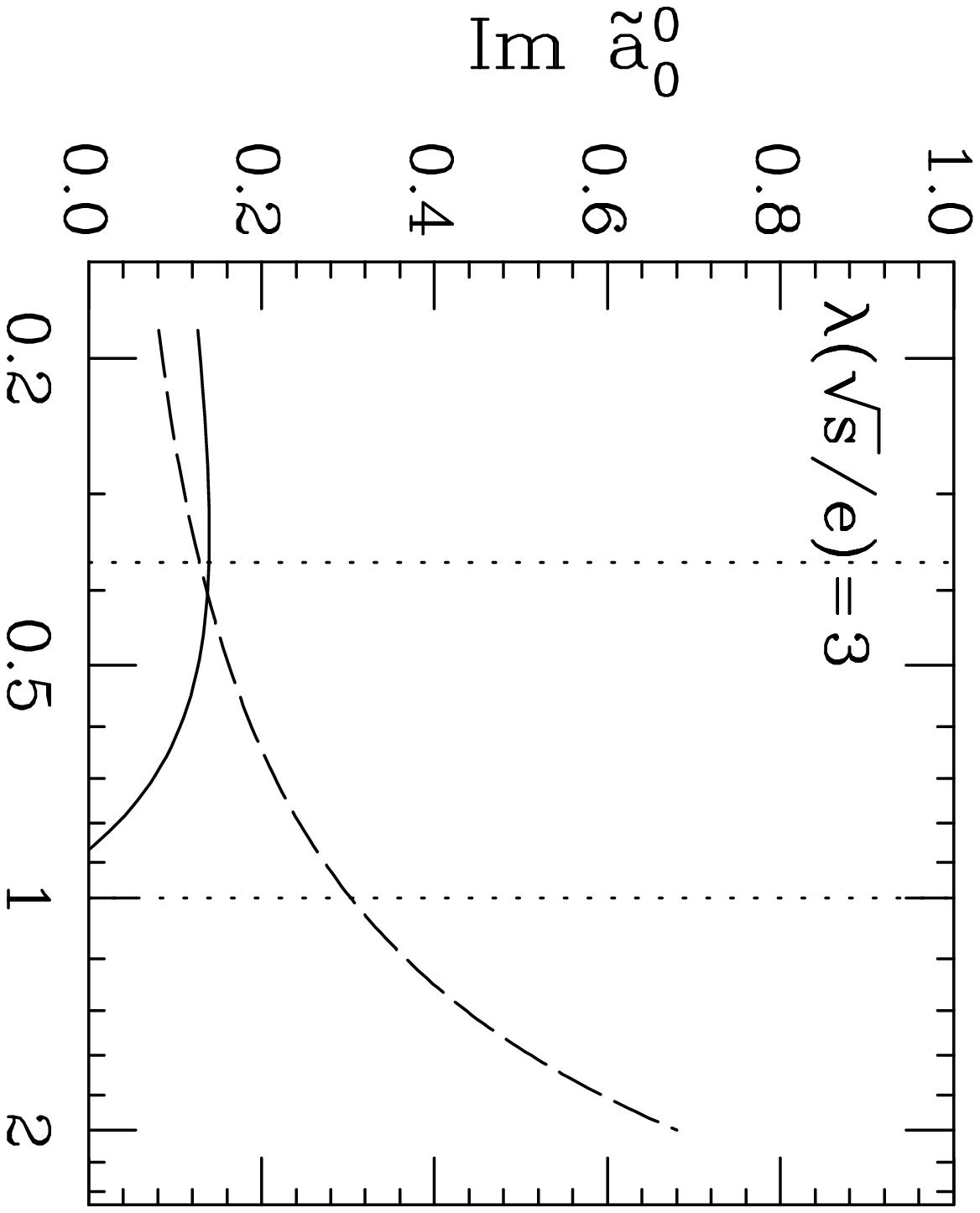}}
}
\centerline{
\epsfysize=2.2in \rotate[l]{\epsffile{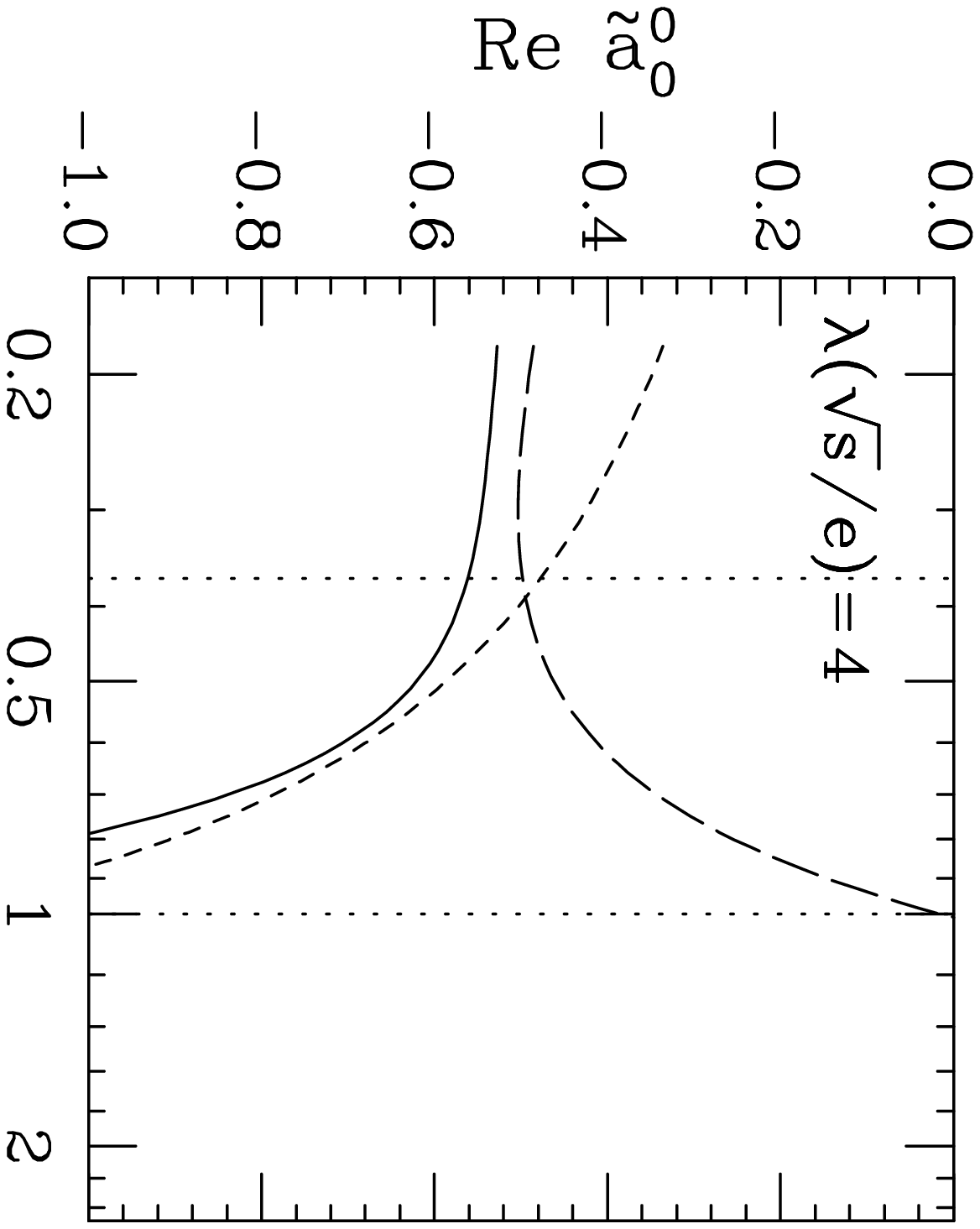}}
\hspace{1cm}\epsfysize=2.2in \rotate[l]{\epsffile{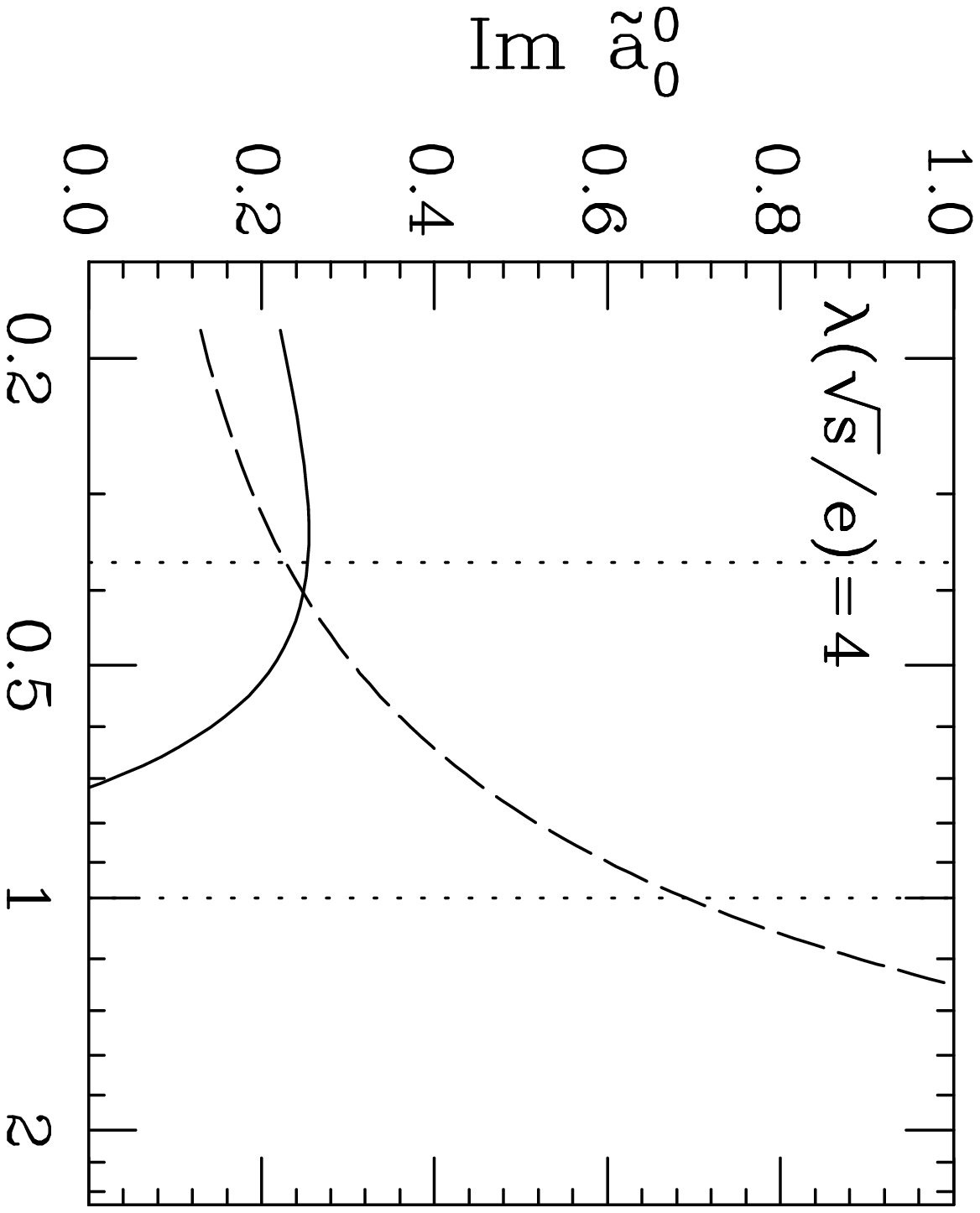}}
}
\centerline{
\epsfysize=2.2in \rotate[l]{\epsffile{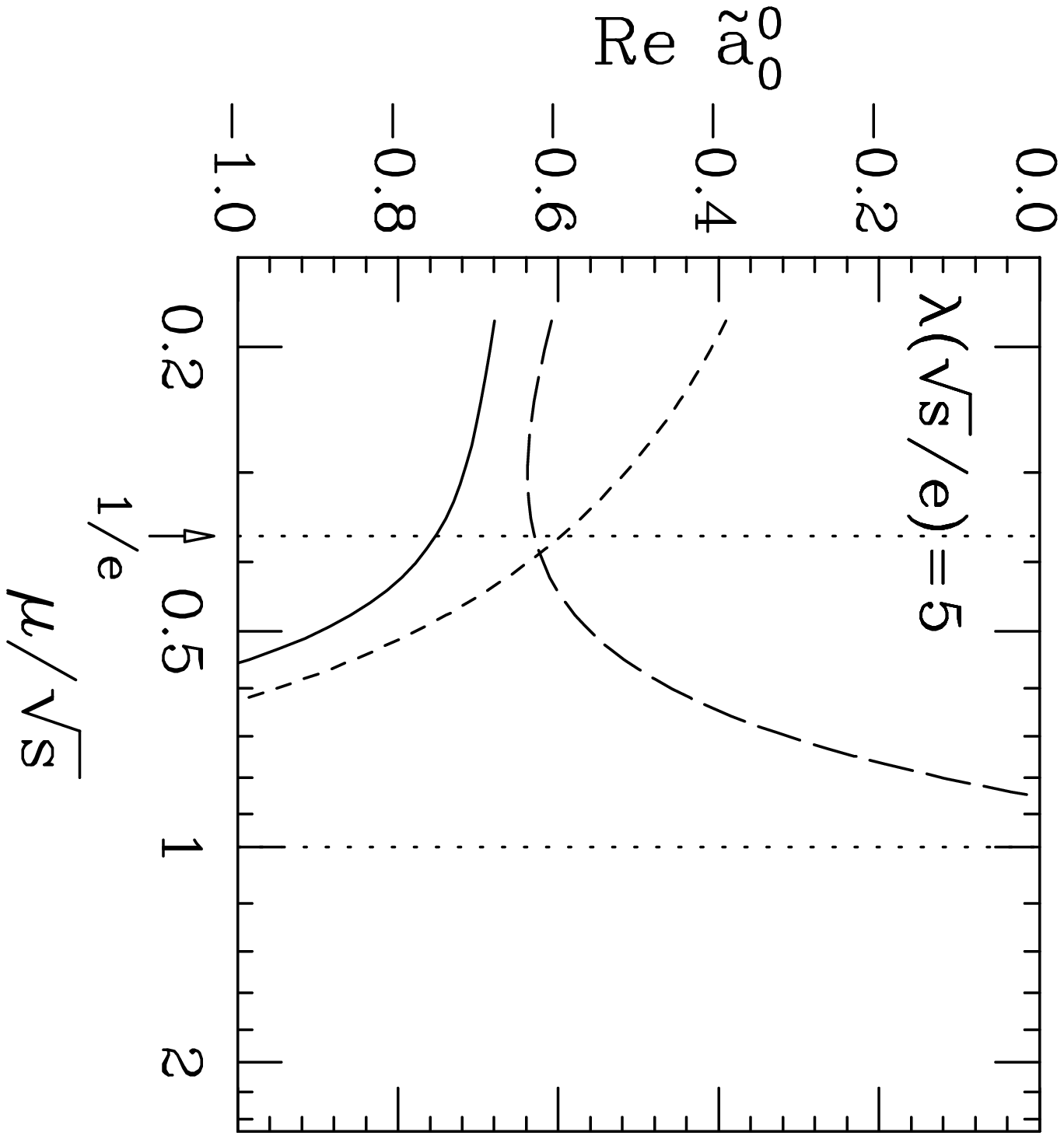}}
\hspace{1cm}\epsfysize=2.2in \rotate[l]{\epsffile{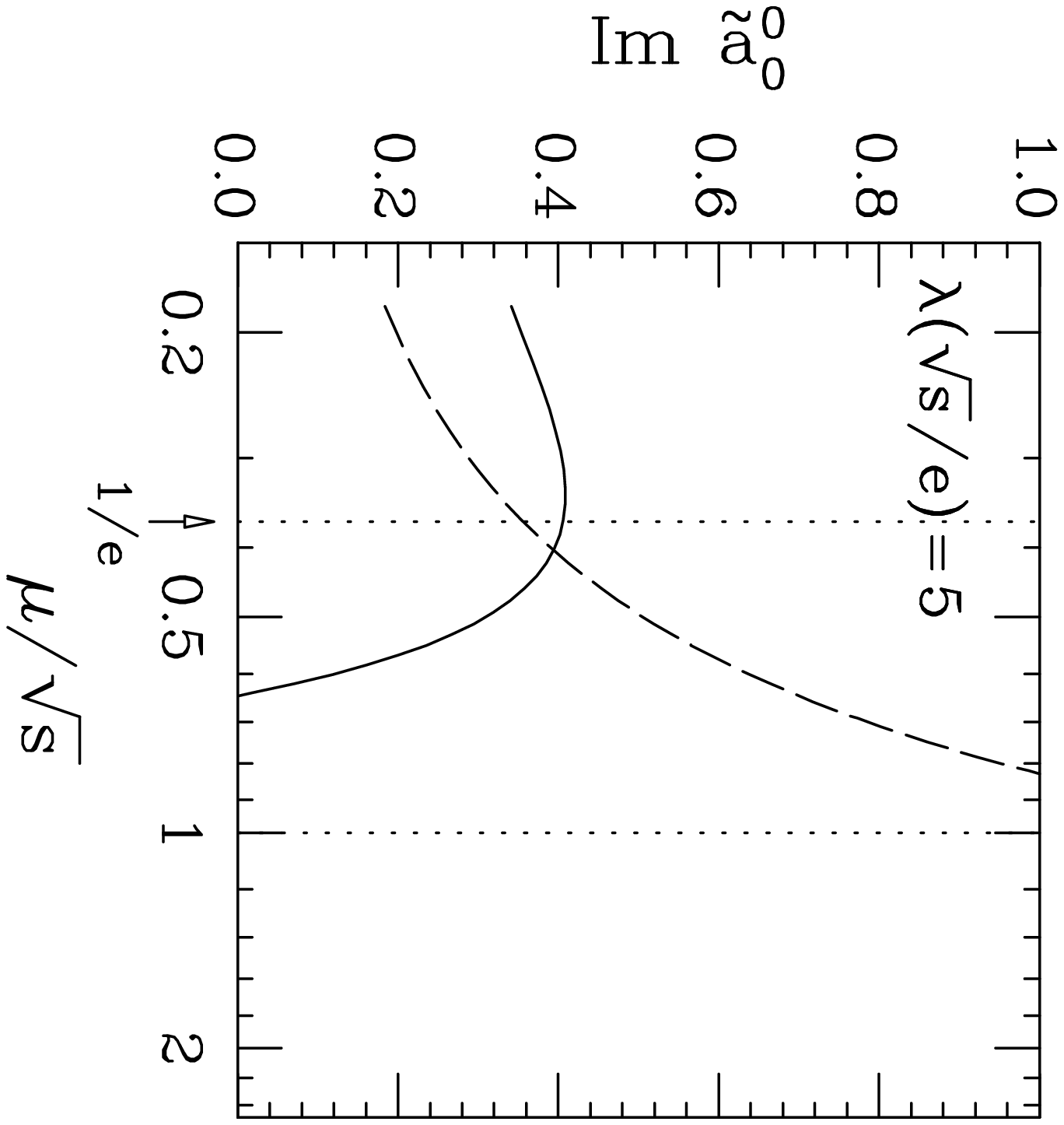}}
}
\vspace{0.2in}
\caption{
  Real part (left column) and imaginary part (right column) of the LL, NLL, and
  NNLL $J=0$ eigenamplitude, $\tilde a_0^0$, for Higgs- and Goldstone-boson
  scattering. The amplitude is shown as a function of $\mu/\protect\sqrt s$,
  fixing $\lambda(\protect\sqrt s/e)$ to be $2.0,\,3.0,\,4.0,$ and 5.0 (from
  top to bottom).  The corresponding values of $\lambda(\protect\sqrt s)$ are
  2.7, 4.6, 7.0, and 9.8.  }
\label{figa0}
\end{figure}

We begin by investigating the size of the radiative corrections and the scale
dependence of the amplitude.  In Fig.~\ref{figa0} we show the real and
imaginary parts of $\tilde a_0^0$ at leading order (LL), next-to-leading order
(NLL), and next-to-next-to-leading order (NNLL), for various values of
$\lambda(\sqrt s/e)$, as a function of the renormalization scale $\mu$ (scaled
by $\sqrt s$).  Table \ref{table1} contains a translation of the values of
$\lambda(\sqrt s/e)$ to the conventional quantity $\lambda(\sqrt s)$, and to
some corresponding pairs of $(m_R,\sqrt s)$. A smaller Higgs mass requires a
larger $\sqrt s$ to obtain a given value of $\lambda(\sqrt{s}/e)$ since the
running coupling must evolve over a larger energy range to achieve the same
magnitude of the coupling.

As is evident from Fig.~\ref{figa0}, the size of the radiative corrections is
greatly reduced for the scale $\mu=\sqrt s/e$ in comparison with the scale
$\mu=\sqrt s$.  Furthermore, the scale dependence is much less when the scale
is varied about $\mu=\sqrt s/e$ rather than $\mu=\sqrt s$.  These observations
apply to both the real and imaginary parts of the amplitude.  They support our
finding that the appropriate scale for Higgs- and Goldstone-boson scattering,
at energies large compared with the Higgs mass, is $\mu=\sqrt s/e$.  Judging
from the scale dependence, it seems that perturbation theory begins to break
down around $\lambda(\sqrt s/e)=4.0$.  Note, however, that even for this value
of $\lambda$ the magnitude of the radiative corrections is not very large, so
the size of the corrections does not appear to be a good indication of the
reliability of perturbation theory.

\begin{figure}[tb]
\vspace*{13pt}
\centerline{
\epsfysize=2.5in \rotate[l]{\epsffile{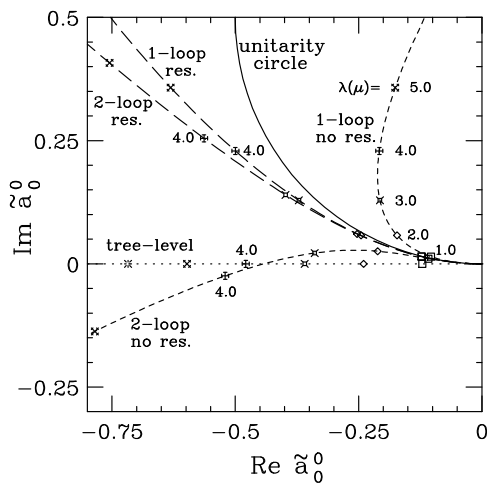}}
}
\vspace{0.2in}
\caption{
  The real and imaginary part of the eigenamplitude $\tilde a_0^0$ plotted in
  an Argand diagram. Shown are the NLL and NNLL results with (long dashes) and
  without (short dashes) summation of the constant 2. The curves are
  parameterized as a function of the running coupling $\lambda(\mu)$, so the LL
  results (dotted curve) coincide in the two approaches. }
\label{figunit}
\end{figure}

A third method of judging the convergence of perturbation theory is to check
the nonperturbative requirement that the eigenamplitude must lie in or on the
unitarity circle.  Plotting an Argand diagram, we show in Fig.~\ref{figunit}
the value of the one-loop and two-loop RG improved $s$-wave eigenamplitude
$\tilde a_0^0$ when taking $\mu=\sqrt{s}/e$ (see Eq.~(\ref{wwwwour})),
indicating various values of the coupling $\lambda(\sqrt s/e)$ (long dashed
curves).  Also shown is the eigenamplitude when taking $\mu=\sqrt{s}$ (see
Eq.~(\ref{wwwwour})) \cite{DJL,DMR2}, indicating various values of
$\lambda(\sqrt s)$ (short dashed curves).\footnote{The values in the case
  $\mu=\sqrt{s}$ are not identical to those in Refs.~\cite{DJL,DMR2} since a
  slightly different initial condition was used for the renormalization-group
  equation in \cite{DJL,DMR2} than is used here.  That initial condition leads
  to the amplitude straying slightly further from the unitarity circle for a
  given value of $\lambda(\sqrt{s})$.}
At leading-order the two approaches coincide (dotted curve) since the choice of
$\mu$ has no influence on the tree-level coefficient.  The fact that, for the
same value of $\lambda$, the amplitudes with the scale $\mu=\sqrt s/e$ lie
closer to the unitarity circle is another way of demonstrating the improved
convergence of perturbation theory with this scale.  We may also use this plot
to again estimate the value of the coupling at which perturbation theory
becomes unreliable.  The next-to-next-to-leading-order amplitude begins to
stray uncomfortably far from the unitarity circle for $\lambda(\sqrt
s/e)\approx 4$. This yields a perturbative upper bound on the running coupling
which is in agreement with our findings above.

Previous analyses, using the scale $\mu = \sqrt s$, concluded that perturbation
theory becomes unreliable for $\lambda(\sqrt s) = 2.0-2.3$
\cite{DJL,DMR2,R,NR}.  This corresponds to $\lambda(\sqrt s/e)=1.6-1.8$.
Figs.~\ref{figa0} and \ref{figunit} suggest that perturbation theory is very
convergent for this range of $\lambda(\sqrt s/e)$.  Choosing $\mu=\sqrt{s}/e$
we conclude that both scale dependence and unitarity indicate that perturbation
theory becomes unreliable for $\lambda(\sqrt s/e)\approx 4$.  This value is
comparable to the simple tree-level unitarity bound of $\lambda <4\pi/3\approx
4.2$ based on $|{\rm Re}\: a_0^0|<1/2$ \cite{DM,LQT,LW1,MVW}.

\subsection{The absence of a strongly-interacting Higgs sector 
at high energies} 

We now ascertain the largest value of $\lambda(\sqrt s/e)$ attainable with the
constraint $\sqrt s < \Lambda$ in order to answer the question posed in the
Introduction: Can the running coupling be strong for energies $\sqrt{s}$ in the
range $m_R < \sqrt{s} < \Lambda$?  It is impossible to define the cutoff
$\Lambda$ precisely, but L\"uscher and Weisz have argued that the effective
lattice cutoff lies roughly between $a^{-1}$ and $2a^{-1}$, by studying the
cutoff effects on Goldstone-boson scattering above the Higgs mass
\cite{LW2}.\footnote{In Ref.~\cite{LW1,LW2}, $a^{-1}$ is denoted by $\Lambda$.
  We use $\Lambda$ to denote the cutoff, which is only proportional to
  $a^{-1}$.} 
The cutoff effects on Goldstone-boson scattering at $2a^{-1}$ are
on the order of ten percent \cite{GKNZ,LW2}.  The authors found that the
relationship between the lattice spacing $a$ and the renormalized coupling
$\hat\lambda(m_R)\equiv \lambda(m_R)/16\pi^2$ is given approximately by the
semi-perturbative two-loop formula
\begin{equation}
\ln\frac{1}{m_Ra} = \frac{1}{\beta_0\hat\lambda(m_R)} + 
\frac{\beta_1}{\beta_0^2}\ln[\beta_0\hat\lambda(m_R)]-\ln C
\label{cutoff}
\end{equation}
where $\beta_0=24$, $\beta_1=-312$ are the one- and two-loop beta-function
coefficients, and $\ln C=1.9$ is a constant which has been obtained
nonperturbatively \cite{LW1,LW2}. (Ref.~\cite{GKNZ} finds a slightly
smaller value, $\ln C=1.445$.)  The consistent solution of the two-loop
renormalization-group equation for $\hat\lambda(\mu)\equiv
\lambda(\mu)/16\pi^2$ is \cite{NR}
\begin{equation}
\frac{\hat\lambda(m_R)}{\hat\lambda(\mu)}=1
-\beta_0\hat\lambda(m_R)\ln\frac{\mu}{m_R}
+\frac{\beta_1}{\beta_0}\hat\lambda(m_R)\ln\frac{\hat\lambda(m_R)}
{\hat\lambda(\mu)}\;.
\end{equation}
Combining these two equations, we obtain an implicit relation between 
$\hat\lambda(\mu)$ and $a$:
\begin{equation}
\frac{1}{\hat\lambda(\mu)}=
-\frac{\beta_1}{\beta_0}\ln[\beta_0\hat\lambda(\mu)]
+\beta_0(C-\ln(\mu a))\;.
\label{A}
\end{equation}

For Higgs- and Goldstone-boson scattering at $\sqrt s=a^{-1}$, our scale
corresponds to $\mu =a^{-1}/e$; for $\sqrt s=2a^{-1}$ (the approximate upper
bound on the cutoff), it corresponds to $\mu = 2a^{-1}/e$.  Solving for the
coupling using Eq.~(\ref{A}), we find $\lambda(a^{-1}/e)=2.7$, and
$\lambda(2a^{-1}/e)=3.5$.  These values are comfortably below the value
$\lambda(\sqrt s/e)=4.0$ at which perturbation theory becomes unreliable.

We therefore conclude that the standard Higgs model is not strongly interacting
at energies above the Higgs mass but below the cutoff.  This is a new result,
and complements the result that the Higgs model cannot be strongly interacting
at energies of the order of the Higgs mass \cite{LW1,LW2}.

\section{Higgs decay}

\indent\indent We now consider the decay amplitude of a heavy Higgs boson.
This is another process which has been calculated at next-to-next-to-leading
order \cite{G,FKKR}, and therefore can also be used to explore the convergence
of perturbation theory at large coupling.  In contrast to Higgs- and
Goldstone-boson scattering, however, the appropriate scale for the running
coupling is the Higgs mass, $\mu=m_R$, not $m_R/e$. Since the energy entering
the decay amplitude is the Higgs mass, one cannot neglect the Higgs mass in the
loop diagrams. Thus the logarithms, which come predominantly from loops
containing Higgs bosons, are not accompanied by the universal constant 2
associated with massless bubble diagrams, in contrast to the case of
high-energy scattering processes.  Recall that this is the same reason the
logarithms in the scattering counterterms are not accompanied by a universal
constant.

Let us consider the maximum allowable Higgs mass, since this yields the maximum
value of the coupling.  Traditionally, this has been obtained from lattice
calculations with the requirement that the cutoff be substantially greater than
the Higgs mass.  Recall that, for example, the upper bound on the coupling of
$\lambda(m_R)=3.2$ (requiring $a^{-1} > 2m_R$) obtained by L\"uscher and Weisz
translates into an upper bound of 630 GeV \cite{LW1}.  The analysis of
Ref.~\cite{GKNZ} yields a similar bound of 680 GeV, using the same lattice
action.  We consider $m_R=700$ GeV, which corresponds to a coupling
$\lambda(m_R)=4.0$.  Based on our experience with Higgs- and Goldstone-boson
scattering, we expect this value to lie just within the perturbative regime.

\begin{figure}[tb]
\vspace*{13pt}
\centerline{
\epsfysize=2.2in \rotate[l]{\epsffile{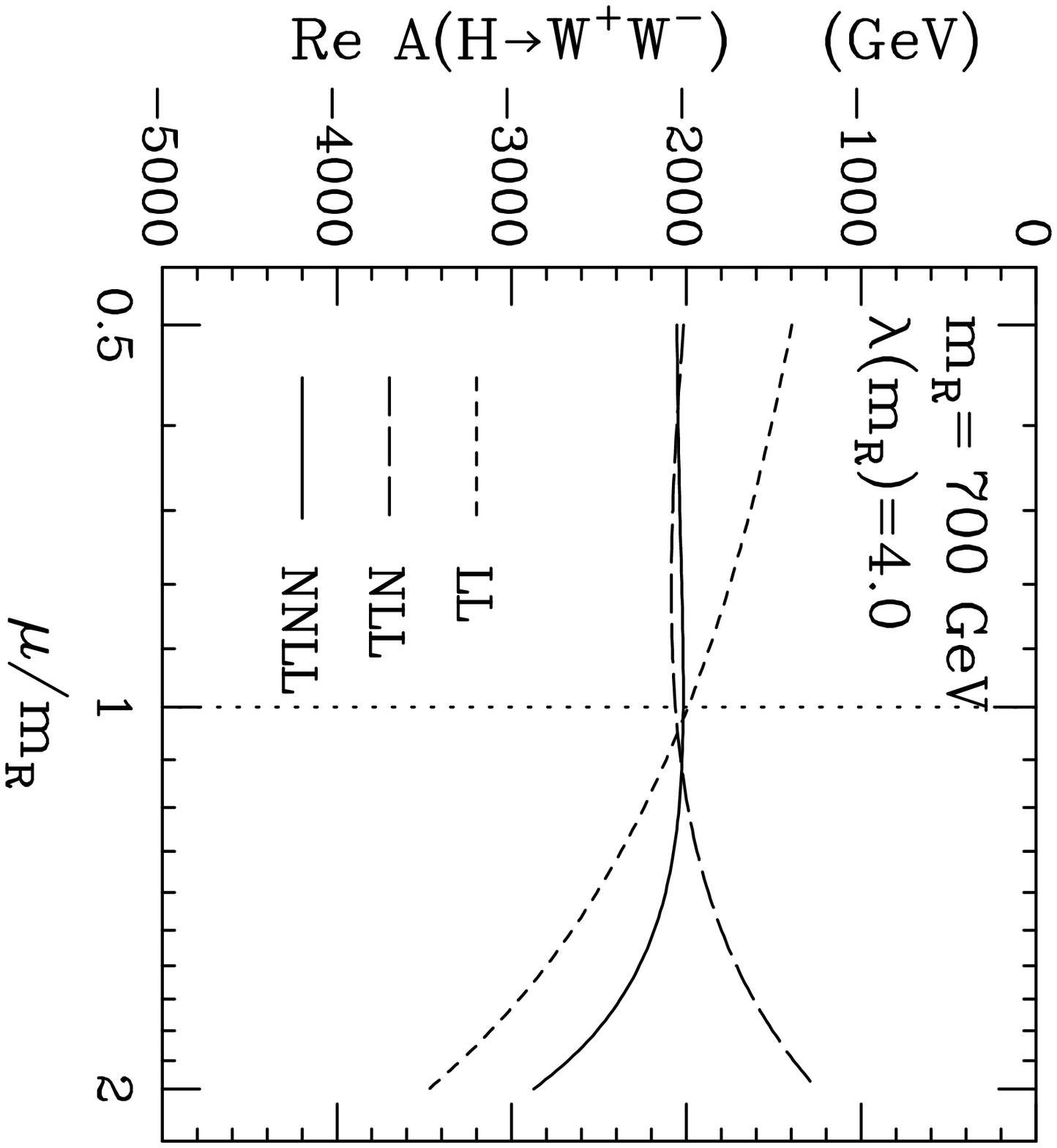}}
\hspace{1cm}\epsfysize=2.2in \rotate[l]{\epsffile{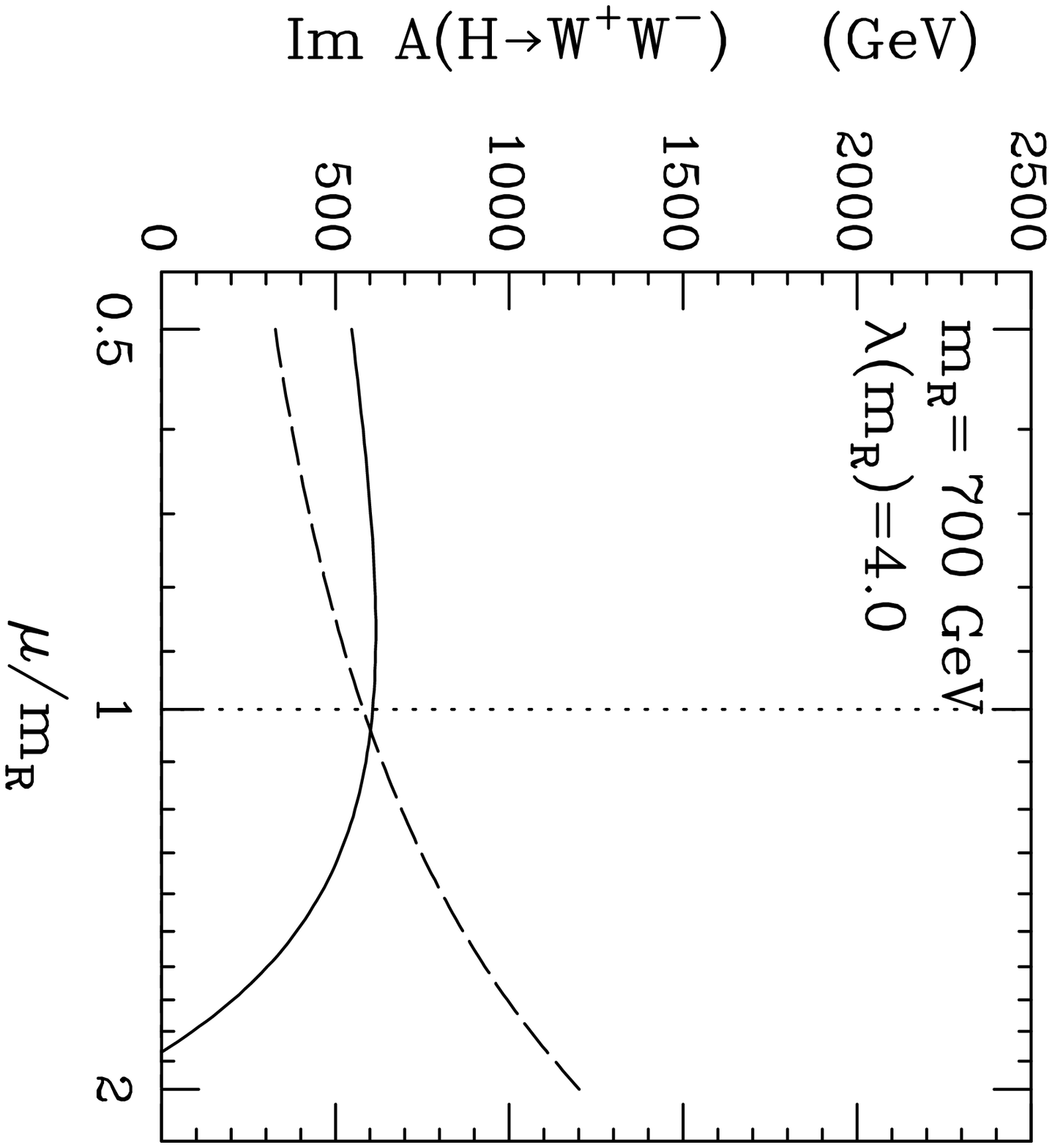}}
}
\vspace{0.2in}
\centerline{
\epsfysize=2.2in \rotate[l]{\epsffile{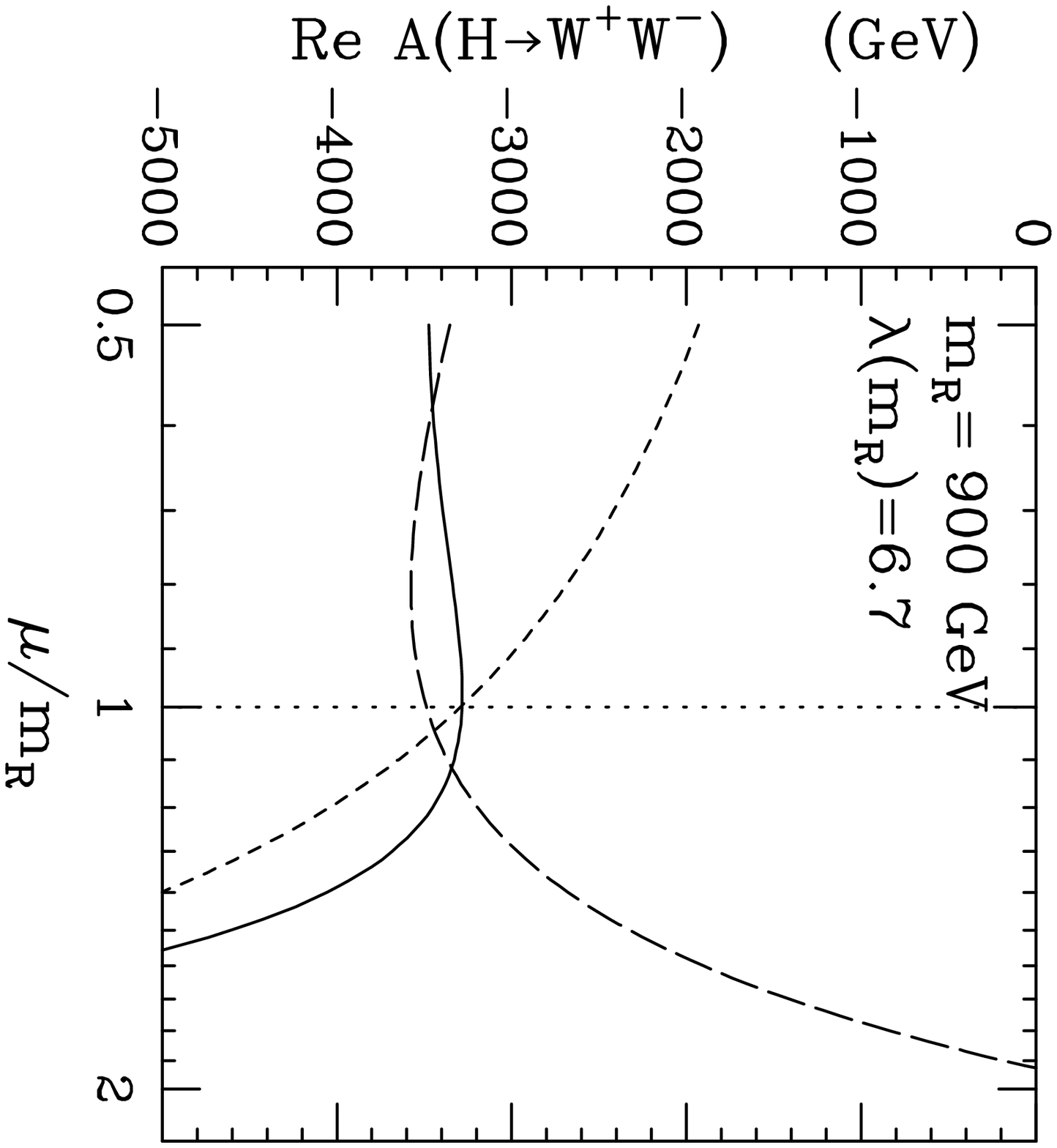}}
\hspace{1cm}\epsfysize=2.2in \rotate[l]{\epsffile{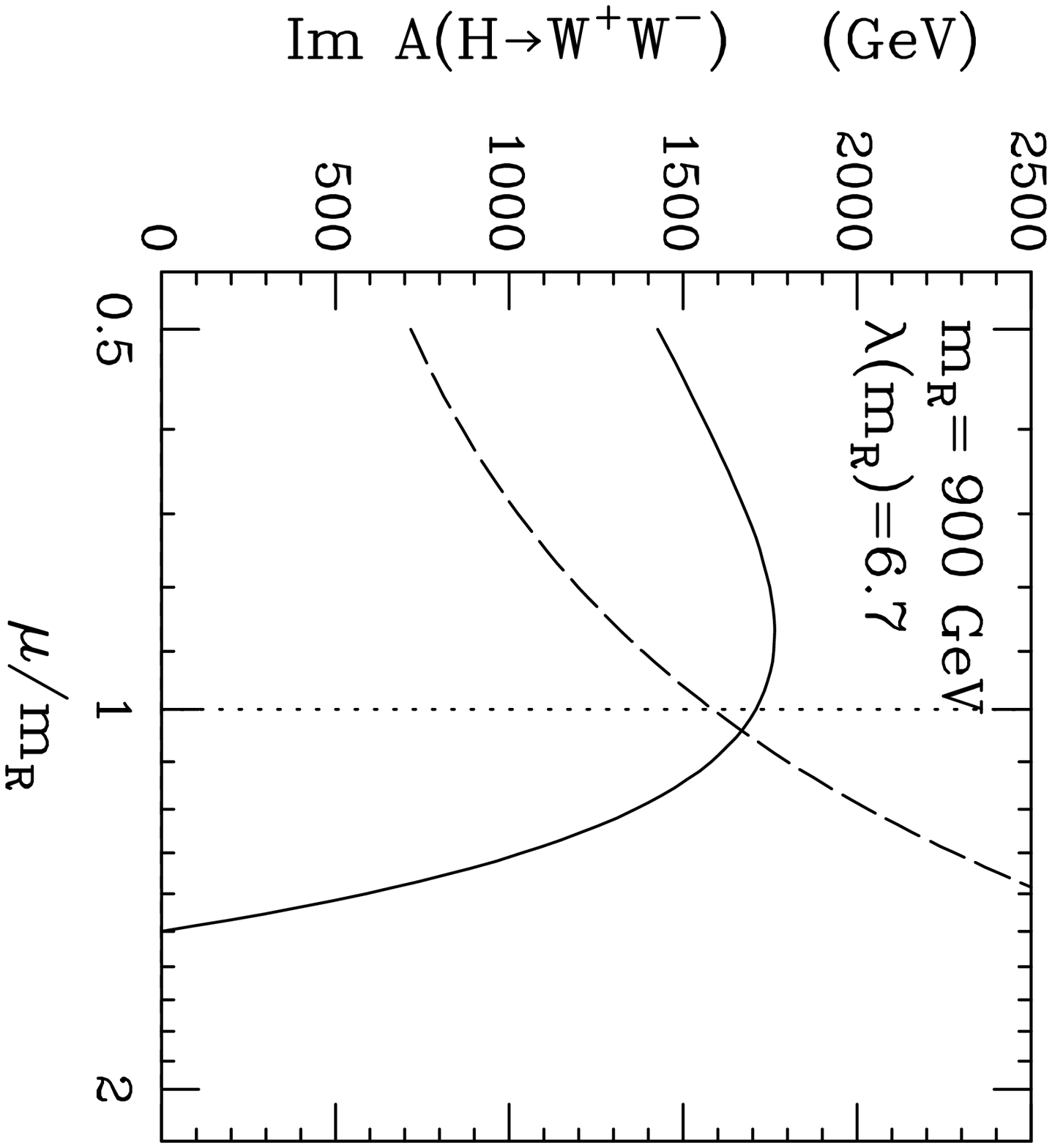}}
}
\vspace{0.2in}
\caption{
  Real and imaginary parts of the leading order, next-to-leading order, and
  next-to-\-next-to-leading order amplitude for Higgs decay to a pair of
  Goldstone bosons as a function of $\mu/m_R$, for $m_R=700$ and 900 GeV.
  }
\label{fighww}
\end{figure}

Written in terms of the running coupling $\lambda(\mu)$, the decay amplitude to
a pair of Goldstone bosons \cite{G,FKKR} becomes \cite{NR}:
\begin{eqnarray}
A(H\to ww) & = & -2v\lambda(\mu)\left[1+\frac{\lambda(\mu)}{16\pi^2}
\left[12\ln\frac{m_R^2}{\mu^2}+1.40-3.61i\pi\right]\nonumber\right.\\
& + &\left.\left(\frac{\lambda(\mu)}{16\pi^2}\right)^2
\left[144\ln^2\frac{m_R^2}{\mu^2}
+(-122.4-86.73i\pi)\ln\frac{m_R^2}{\mu^2}
-34.35-21.00i\right]\right]\;.\nonumber\\
&&
\end{eqnarray}
We show in Fig.~\ref{fighww} the real and imaginary parts of the leading-order,
next-to-leading order, and next-to-next-to-leading-order decay amplitude for
$m_R=700$ and 900 GeV, as a function of $\mu/m_R$.  In the case of 700 GeV (top
figures), the amplitude is rather insensitive to the scale in the vicinity of
$\mu=m_R$, while it is rather sensitive to the scale above and below this
region.  This supports our statement that the appropriate scale for the Higgs
decay amplitude is indeed the Higgs mass.  The sensitivity of the amplitude to
the scale decreases with increasing order in perturbation theory for $\mu=m_R$,
indicating that perturbation theory is reliable.  Given the size of the
coupling, the corrections to the decay amplitude are remarkably small, a
feature we also observed in the case of Higgs- and Goldstone-boson scattering.
The case of 900 GeV (bottom figures in Fig.~\ref{fighww}) corresponds to a
coupling $\lambda(m_R)=6.7$, which is quite large.  The scale dependence of the
amplitude has significantly increased when compared with the case of 700 GeV.
The numerical studies of \cite{NR}, which investigate the scale dependence of
the decay {\it width} rather than the amplitude, find the perturbative approach
to be unreliable for $m_R\gtrsim 700$ GeV. All these findings confirm that the
maximal value of $m_R$ found in lattice studies is within the perturbative
range, supporting the original work of L\"uscher and Weisz.

In Ref.~\cite{HKNV,HNV} it is speculated that perturbation theory seriously
underestimates the Higgs width for large Higgs mass, $m_R \approx 700$ GeV.
This is based on a calculation of the Higgs width in the $1/N$ expansion.  It
is difficult to reconcile this with the fact that perturbation theory is
apparently reliable for such a Higgs mass as seen in recent two-loop
calculations \cite{G,FKKR}.  The discrepancy with the $1/N$ calculation
disappears when the Goldstone bosons are given a significant mass.
The Higgs width on the lattice \cite{GKWZ} (which is
calculated with a significant Goldstone-boson mass $\sim m_R/3$) also suggests
agreement with perturbation theory for a Higgs mass of roughly 700 GeV
\cite{FKKR}.  An extrapolation of the lattice results to the limit of (nearly)
massless Goldstone bosons, as in the perturbative
calculations, would be of interest.

\section{Phenomenological implications}

If and when a Higgs boson is discovered, it will be interesting to measure
vector-boson scattering at energies above the Higgs resonance. In the standard
model, the
Higgs boson is responsible for regulating the growth of the amplitude for
longitudinal-vector-boson scattering with energy: At low energies, the
scattering amplitude is proportional to $s/v^2$; above the resonance, it is
proportional to $\lambda$.  Observing this behaviour experimentally will be
challenging.

As an example we look at the effect of our summation procedure on the case of
high-energy $W^+W^-\rightarrow ZZ$ scattering.  This process is of interest for
future colliders such as the LHC or linear $e^+e^-$ and $\mu^+\mu^-$ 
colliders.  Its
amplitude is immediately given by the generic amplitude $a(s,t,u)$; recall
Eq.~(\ref{wwzz1}).  Using the two-loop result of \cite{DMR2}, the NNLL cross
section with its explicit $\mu$ dependence is\cite{NR}
\begin{eqnarray}
\sigma(s)\, 
& = & \frac{1}{8\pi s} [\lambda(\mu)]^2\,
\Biggl[\,1\,+
\left( 24 \ln \frac{s}{\mu^2} -
       \, 42.65 \right) \,\frac{\lambda(\mu)}{16\pi^2}\,
\nonumber \\ 
&& \phantom{\frac{1}{8\pi s} [\lambda(\mu)]^2\, }
+ \left( 432 \ln^2 \frac{s}{\mu^2} - 1823.3\ln \frac{s}{\mu^2} 
%   + 24.0\ln \frac{s}{m_R^2} 
   +\;2457.9\,\right) \frac{\lambda^2(\mu)}{(16\pi^2)^2}\,
+\; {\rm O}\left(\lambda^3(\mu)\right)\,\Biggr]\,,
\label{wwzzcross}
\end{eqnarray}
where the anomalous dimension prefactor has been neglected since it is close to
unity for the values of $\sqrt s$ and $m_R$ considered here.  Thus the product
$s\sigma$ depends only on the three-loop running coupling and the ratio
$\mu/\sqrt s$.

\begin{figure}[tb]
\vspace*{13pt}
\centerline{
\epsfysize=2.5in \rotate[l]{\epsffile{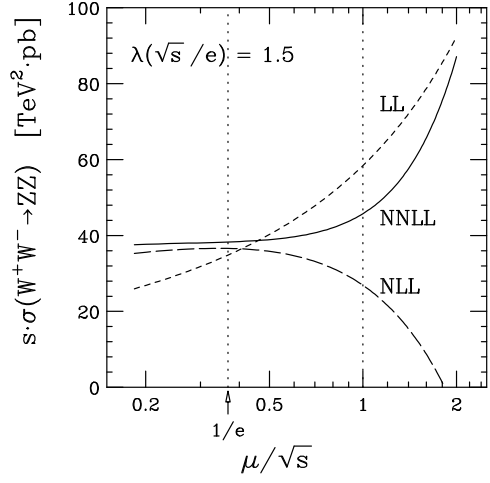}}
}
\vspace{0.2in}
\caption{
  The $\mu$-dependence of the scaled cross section of $W_L^+W_L^-\rightarrow
  Z_LZ_L$ for $\lambda(\protect\sqrt{s}/e)=1.5$ in the high-energy
  approximation. Gauge and Yukawa coupling contributions are neglected.  }
\label{figcross}
\end{figure}

In Fig.~\ref{figcross} we show $s\sigma$ as a function of $\mu/\sqrt s$, fixing
the running coupling such that $\lambda(\mu=\sqrt s/e)$ is equal to 1.5.  Using
the scale $\mu=\sqrt s/e$, we find the size of the radiative corrections to be
significantly reduced.  In addition, the reduced scale dependence around
$\mu=\sqrt{s}/e$ is clearly visible.  The leading-log approximation with the
conventional scale $\mu=\sqrt s$ overestimates the magnitude of the cross
section by more than 30\%, whereas the scale $\mu=\sqrt s/e$ yields a
leading-log result only slightly less than the NLL and NNLL results.  We
conclude that phenomenological studies based on tree-level results are much
more reliable when using $\mu=\sqrt s/e$.

\begin{figure}[tb]
\vspace*{13pt}
\centerline{
\epsfysize=2.5in \rotate[l]{\epsffile{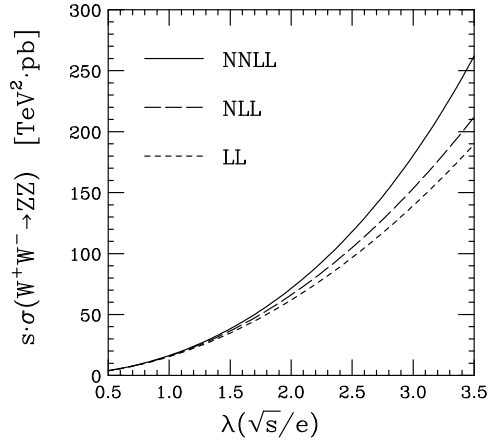}}
}
\vspace{0.2in}
\caption{
  The scaled cross section of $W_L^+W_L^-\rightarrow Z_LZ_L$ for
  $0.5<\lambda<3.5$ in the high-energy approximation, fixing
  $\mu=\protect\sqrt{s}/e$. Gauge and Yukawa coupling contributions are
  neglected.  The values of $\lambda(\protect\sqrt{s}/e)$ can be converted to
  $(m_R,\protect\sqrt{s})$ using Table \protect\ref{table1}. }
\label{figcross2}
\end{figure}

Fixing $\mu=\sqrt{s}/e$, we show in Fig.~\ref{figcross2} the LL, NLL, and NNLL
result for $s\sigma$ as a function of $\lambda(\sqrt{s}/e)$, displaying the
perturbative range $0.5<\lambda(\sqrt{s}/e)<3.5$.  Using Table I, the value of
the running coupling can be related to the desired Higgs mass and the
center-of-mass energy of the incoming $W_LW_L$ pair.  Standard analyses of the
cross section using $\mu=\sqrt{s}$ lead to large uncertainties \cite{R} for
running coupling larger than about 2.  The improved scale greatly reduces the
one-loop and two-loop corrections, allowing for predictive cross sections even
for $\lambda(\sqrt{s}/e)$ close to 4.

The high-energy amplitude given here is completely based on the four-point
interactions of the Higgs sector, a good approximation for $\sqrt s> 2-3m_R$
\cite{R}. For smaller values of $\sqrt s$, the three-point interactions
dominate over the four-point coupling.  In addition, the electroweak gauge
couplings contribute to the cross section, making a measurement of the Higgs
coupling difficult.

\section{Conclusions}

\indent\indent In this paper we resolve the mystery, raised in Ref.~\cite{DJL}
and deepened in Refs.~\cite{DMR2,R,NR}, that the perturbative calculation of
Higgs- and Goldstone-boson scattering, at energies large compared with the
Higgs mass, is apparently unreliable for rather small values of the running
coupling, $\lambda = 2.0 - 2.3$.  The resolution lies in the choice of the
scale in the running coupling $\lambda(\mu)$.  All previous analyses have
implicitly or explicitly used $\mu=\sqrt s$.  We argue that a more appropriate
scale is $\mu=\sqrt s/e$, and show that this scale leads to a dramatic
improvement in the convergence of perturbation theory for Higgs- and
Goldstone-boson scattering.  We find that perturbation theory is apparently
reliable up to a coupling $\lambda\approx4$, consistent with the perturbative
unitarity bound of $\lambda<4\pi/3\approx 4.2$.

With the improved perturbation theory, we address the question of whether
Higgs- and Goldstone-boson scattering can become strongly interacting at
energies above the Higgs mass but below the cutoff, modeled by the inverse
lattice spacing.  We find that the value of the coupling for Higgs- and
Goldstone-boson scattering at the cutoff is within the perturbative domain: a
strongly-interacting standard Higgs model at high energies is excluded.  This 
is a
new result, and complements the result that the Higgs sector cannot be strongly
interacting at energies near the Higgs mass \cite{LW1,LW2}.

We also consider the decay amplitude of the Higgs boson to Goldstone bosons.
In this case we argue that the appropriate scale is the Higgs mass, and we show
that perturbation theory is apparently reliable up to a coupling of
$\lambda(m_R)=4.0$, which corresponds to a Higgs mass of 700 GeV.  This
supports the conclusions of Ref.~\cite{LW1,LW2}.  
This is difficult to reconcile with the observation, made in Ref.~\cite{HKNV},
that
there is a discrepancy between the Higgs width calculated in the $1/N$
expansion and in perturbation theory for $m_R\approx 700$ GeV.  A lattice
calculation of the Higgs width with an extrapolation to the case of (nearly)
massless Goldstone bosons would be desirable.

The most important aspect of our work is the realization that the apparent
breakdown of perturbation theory at weak coupling is simply due to a poor
choice of scale in the running coupling.  Our argument for the scale $\mu=\sqrt
s/e$ is based on an analysis of the constant which accompanies the logarithm in
the one-loop bubble diagram.  It may be possible to refine this argument
further.  One might be able to develop a scale-fixing scheme analogous to the
BLM method \cite{BLM}: the number of Goldstone bosons, $n_g$, could play the
role of $n_f$, the number of light fermions. (The terms proportional to $n_g$
connected to the counterterms should not be included in a BLM analysis.)
Naively applying the BLM method at one loop leads to the same scale which we
advocate.  It is also interesting that our scale lies in the region where the
amplitude is quite insensitive to the choice of scale. Therefore the principle
of minimal sensitivity \cite{S} is also expected to lead to a scale close to
$\mu=\sqrt{s}/e$.

\acknowledgements

\indent\indent We are grateful for conversations and correspondance with
M.~Beneke, A.~El-Khadra, U.~Heller, P.~Mackenzie, M.~Neubert, U.~Nierste, and
P.~Weisz.  This work was performed in part at the Aspen Center for Physics.
S.~W. was supported in part by Department of Energy grant DE-FG02-91ER40677.

\appendix

\section{Relation of \mbox{$m_R$} to the Higgs boson mass $m_H$}
\label{mass}

The quantity $m_R^2$ is defined as the zero of the real part of the inverse
Higgs-boson propagator.  A physical definition of the Higgs mass, $m_H$, is the
real part of the pole (in the energy plane) of the Higgs propagator
\cite{LW1,VW}.  This definition is process-independent and field-redefinition
invariant.  The relation between $m_R$ and $m_H$ is
\begin{equation}
m_H^2 
= m_R^2\left[1+\frac{\Gamma_H^2}{4}+\cdots\right]
= m_R^2\left[1+\frac{9}{64}\left(\frac{\lambda(m_R)}{4\pi}\right)^2
  +\cdots\right]\;,
\label{MR}
\end{equation}
where $\Gamma_H$ is the Higgs width. The ${\cal O}(\lambda^3)$ term in
Eq.~(\ref{MR}) is given in Ref.~\cite{VW}. For $m_R \le 1200$ GeV, the
distinction between $m_R$ and $m_H$ is numerically negligible, so one may
safely refer to $m_R$ as the physical Higgs mass.

\section{The running coupling $\lambda(\mu)$ and the $\beta$ function to three
  loops}
\label{runcoup}

To obtain renormalization-group-improved scattering amplitudes, the evolution
of $\lambda(\mu)$ as a function of $\mu$ is needed. It is dictated by the
renormalization group equation,
\begin{equation}
\label{rge}
\frac{d\lambda(\mu)}{d\ln\mu} = \beta(\lambda(\mu))\;,
\end{equation}
with the initial condition imposed by Eq.~(\ref{COUPLING}).  For large values
of $\lambda$ we can neglect all gauge and Yukawa coupling contributions to the
beta function. The three-loop result is \cite{LW2,NR}:
\begin{equation}
\beta(\lambda)
=24\frac{\lambda^2}{16\pi^2} 
\left[ 1 - 13\frac{\lambda}{16\pi^2}
         + 176.6\left(\frac{\lambda}{16\pi^2}\right)^2
\right]\;.
\label{beta}
\end{equation}
Neglecting the appropriate powers of $\lambda$, these equations determine the
$n$-loop running coupling for $n\leq3$. Explicitly, the one-loop running
coupling is
\begin{eqnarray}
\lambda(\mu)&=&\frac{\lambda(m_R)}
{1 - 12\,\frac{\lambda(m_R)}{16\pi^2}\ln\left(\frac{\mu^2}{m_R^2}\right) }\,.
\end{eqnarray}
At higher order the solution of Eq. (\ref{rge}) is not unique anymore, and
various solutions are discussed in \cite{NR}. We take the ``consistent
solution'' introduced in \cite{NR}.  The $n$-loop running coupling sums the
leading logs, next-to-leading logs, and next-to-next-to-leading logs of the
physical amplitudes for $n=1,2,3$, respectively.

\section{Nonzero Higgs mass effects in the bubble diagram}
\label{massbub}

The Feynman amplitude for Higgs- and Goldstone-boson scattering receives
contributions from the {\it massless} scalar bubble diagram,
\begin{equation}
B(p^2)\equiv B_0(p^2;m_1^2=0,m_2^2=0)=\frac{1}{\epsilon} - \gamma +
        \ln\left(\frac{4\pi\mu^2}{-p^2}\right) + 2 + O(\epsilon)
\end{equation}
where $p$ is the incoming four-momentum, and $m_1,m_2$ are the internal
particle masses. The logarithm is accompanied by the constant 2.  In the limit
$p^2\gg m_R^2$, the bubble diagrams with internal Higgs propagators
($m_1^2=m_2^2=m_R^2$) are also well approximated by the massless case $B(p^2)$.
The counterterms appearing in Eq.~(\ref{ampstrt}) receive contributions from
bubble diagrams with $p^2=0$ or $m_R^2$: the masses of the internal Higgs
bosons cannot be neglected, and the corresponding finite pieces are different
from 2.  To illustrate this we list the different bubble contributions occuring
at one loop:
\begin{eqnarray}
B_0(m^2;0,0) &=& \frac{1}{\epsilon} - \gamma +
        \ln\left(\frac{4\pi\mu^2}{m^2}\right) + 2 + i\pi + O(\epsilon)\,,\\
B_0(m^2;m^2,m^2) &=& \frac{1}{\epsilon} - \gamma +
        \ln\left(\frac{4\pi\mu^2}{m^2}\right) + 2 - \frac{\pi}{\sqrt{3}} 
        + O(\epsilon)\,,\\
B_0(0;0,m^2) &=& \frac{1}{\epsilon} - \gamma +
        \ln\left(\frac{4\pi\mu^2}{m^2}\right) + 1 + O(\epsilon)\,,
\end{eqnarray}

\section{The two-loop scattering graphs}
\label{scatt}
At two loops, the only relevant high-energy scattering topologies
are $A(p^2)$ and $[B(p^2)]^2$. Their exact results are given in \cite{DMR1}.
Expanding in powers of $\epsilon$ and neglecting $O(\epsilon)$ terms, they are
evaluated as:
\begin{eqnarray}
A(p^2) &=&  \frac{(4\pi e^{-\gamma})^{2\epsilon}}{(4\pi)^4}
\left(\frac{\mu^2}{-p^2}\right)^{2\epsilon}
\left( \frac{1}{2\epsilon^2} + \frac{5}{2\epsilon} +
\frac{19}{2}+\frac{1}{2}\zeta(2) + O(\epsilon)\,\right)
\;,\\  
\left[B(p^2)\right]^2 &=& \frac{(4\pi e^{-\gamma})^{2\epsilon}}{(4\pi)^4}
\left(\frac{\mu^2}{-p^2}\right)^{2\epsilon}
\left( \frac{1}{\epsilon^2} + \frac{4}{\epsilon} +
12-\zeta(2) + O(\epsilon) \right)
\;.
\end{eqnarray}

\section{Summing powers of the bubble diagram}
\label{resum}

The scale $\mu=\sqrt s/e$ is motivated by the summation of the contribution
$2^n$ which comes from the $[B(p^2)]^n$ terms at $n$ loops. Since $B(p^2)$ is
ultraviolet divergent, the $O(\epsilon^{m})$ terms will also contribute at
orders $n>m\geq 0$. Here we show that the improved scale also sums those
contributions partially, at least as checked up to $n=5$ loops.  The exact
result of $B(p^2)$ to all orders in $\epsilon$ is
\begin{eqnarray}
\label{bubexact}
B(p^2) 
&=& \frac{(4\pi e^{-\gamma})^\epsilon}{16\pi^2} 
\left(\frac{\mu^2}{-p^2}\right)^\epsilon  
\frac{1}{\epsilon\, (1-2\epsilon)}\, 
\exp\left\{\sum_{n=2}^{\infty}\frac{\epsilon^n\zeta(n)}{n}
\left[2-2^n+(-1)^n\right]\right\} \,,
\end{eqnarray}
where $\zeta$ is the Riemann Zeta function.  Expanding up to $O(\epsilon^4)$
yields the numerical result
\begin{eqnarray}
B(p^2) 
&=& \frac{(4\pi e^{-\gamma})^\epsilon}{16\pi^2}
\left(\frac{\mu^2}{-p^2}\right)^\epsilon  \times \nonumber \\
& & \left( \frac{1}{\epsilon} + 2 + 3.1775\epsilon
+ 3.5502\epsilon^2  + 3.9212\epsilon^3 + 3.7203\epsilon^4
+ O(\epsilon^5)       \right)     \;.
\end{eqnarray}
The $O(\epsilon^{n-1})$ term of this expansion contributes to the finite part
of the perturbative amplitude at $n$ loops and even higher orders.  Factoring
out the constant 2 which is summed by the scale choice $\mu=\sqrt s/e$, we find
that the coefficients of the power series in $\epsilon$ of the previous
equation are reduced in magnitude:
\begin{eqnarray}
B(p^2) 
&=& \frac{(4\pi e^{-\gamma})^\epsilon}{16\pi^2}
\left(\frac{e^2\mu^2}{-p^2}\right)^\epsilon \times \nonumber \\
& & \left( \frac{1}{\epsilon}  + 0 + 1.1775\epsilon
- 0.1381\epsilon^2  + 1.1757\epsilon^3 - 0.1916\epsilon^4
+ O(\epsilon^5)               \right)    \,.
\end{eqnarray}
It is also possible to completely cancel the coefficients to all orders using
the $G$ scheme\cite{CHE}.

\clearpage

\centerline{\bf TABLES}

\begin{table}[ht]
\caption{
Relating the values of $\lambda(\protect\sqrt s/e)$ to $\lambda(\protect\sqrt
s)$  using the three-loop renormalization-group equation. The initial 
condition on the running coupling is given by Eq.~(\protect\ref{COUPLING}).
Some representative values of corresponding pairs of $(m_R,\protect\sqrt s)$
are also given, requiring $\protect\sqrt s>m_R$.
}   
\label{table1}
\medskip
\begin{displaymath}
  \begin{array}{|c|c|c|c|c|c|c|} 
  \hline 
& \multicolumn{6}{c|}{  \rule{0pt}{14pt}\lambda\,(\mu\!=\!\!\protect\sqrt s/e)
\quad\left[\:\lambda\,(\mu\!=\!\!\protect\sqrt s)\:\right] }  \\
\cline{2-7}
& \rule{0pt}{14pt}\;0.50 \:\:[\,0.54\,]\; &\; 1.0 \:\:[\,1.2\,]\; 
&\; 2.0\:\:[\,2.7\,]\; &\; 3.0\:\: [\,4.6\,]\; &\; 4.0\:\: [\,7.0\,]\; 
&\;  5.0\:\: [\,9.8\,]\; \\ 
  \hline 
\; \rule{0pt}{14pt} m_R \:\:({\rm GeV}) & \multicolumn{6}{c|}{\protect\sqrt s
\quad({\rm GeV})} 
\\ 
  \hline 
    250 & 4.4\:{\rm E}\,02   &  4.6\:{\rm E}\,05  &  1.8\:{\rm E}\,07  &
    6.7\:{\rm E}\,07 &  1.3\:{\rm E}\,08  &  2.1\:{\rm E}\,08   \\  
    300 &       -            &  9.3\:{\rm E}\,03  &  3.6\:{\rm E}\,05  &
    1.3\:{\rm E}\,06 &  2.7\:{\rm E}\,06  &  4.2\:{\rm E}\,06   \\  
    350 &       -            &  8.7\:{\rm E}\,02  &  3.4\:{\rm E}\,04  &
    1.3\:{\rm E}\,05 &  2.6\:{\rm E}\,05  &  4.0\:{\rm E}\,05   \\  
    400 &       -            &          -         &  7.3\:{\rm E}\,03  &
    2.7\:{\rm E}\,04 &  5.5\:{\rm E}\,04  &  8.6\:{\rm E}\,04   \\ 
    500 &       -            &          -         &  1.2\:{\rm E}\,03  &
    4.5\:{\rm E}\,03 &  9.0\:{\rm E}\,03  &  1.4\:{\rm E}\,04   \\  
    600 &       -            &          -         &         -          &
    1.7\:{\rm E}\,03 &  3.4\:{\rm E}\,03  &  5.3\:{\rm E}\,03   \\  
    700 &       -            &          -         &         -          &
    9.2\:{\rm E}\,02 &  1.9\:{\rm E}\,03  &  2.9\:{\rm E}\,03   \\  
    800 &       -            &          -         &         -          &
    -        &  1.3\:{\rm E}\,03  &  2.0\:{\rm E}\,03   \\  
    900 &       -            &          -         &         -          &
    -        &  9.6\:{\rm E}\,02  &  1.5\:{\rm E}\,03   \\  
   1000 &       -            &          -         &         -          &
   -        &         -          &  1.2\:{\rm E}\,03   \\  
  \hline 
  \end{array}
\end{displaymath} 
\end{table}

\newpage

\centerline{\bf FIGURES}
\vspace{3cm}

\end{document}